\documentclass[fleqn,usenatbib]{mnras}

\usepackage{newtxtext,newtxmath}

\usepackage[T1]{fontenc}

\DeclareRobustCommand{\VAN}[3]{#2}
\let\VANthebibliography\thebibliography
\def\thebibliography{\DeclareRobustCommand{\VAN}[3]{##3}\VANthebibliography}

\usepackage{graphicx}	
\usepackage{amsmath}	
\usepackage{bm}
\usepackage[normalem]{ulem}

\newcommand\pc{\mbox{{$\rm pc$}}}

\newcommand\K{\mbox{{$\rm K$}}}
\newcommand\Msun{\mbox{$\rm M_{\sun}$}}
\newcommand\cc{\mbox{$\rm cm^{-3}$}}

\newcommand\myr{\mbox{$\rm Myr$}}

\newcommand\sfrunit{\mbox{$\rm M_{\sun}\,yr^{-1}$}}

\newcommand\HII{{\mbox{H{\scriptsize \sc II}}}}

\title[Clump-based SF model for galaxy simulations]{A new clump-based star formation model for galaxy simulations: implications for high-redshift compact star clusters} 

\author[S. Horie and H. Yajima]{
Shu Horie\thanks{E-mail: shorie@ccs.tsukuba.ac.jp (SH)} and
Hidenobu Yajima
\\
Center for Computational Sciences, University of Tsukuba, Ten-nodai, 1-1-1 Tsukuba, Ibaraki 305-8577, Japan\\
}

\date{Accepted XXX. Received YYY; in original form ZZZ}

\pubyear{\the\year{}}

\begin{document}
\label{firstpage}
\pagerange{\pageref{firstpage}--\pageref{lastpage}}
\maketitle

\begin{abstract}
We develop a new star formation model for galaxy simulations in which star-forming gas clumps are identified on-the-fly and converted into stars with an efficiency $\epsilon_{\rm SF}$ set by the clump surface density $\Sigma_{\rm c}$, calibrated against radiation hydrodynamics simulations of star cluster formation.
Applying this model to isolated disc galaxies embedded in haloes of $M_{\rm h}=10^{9}$, $10^{10}$ and $10^{11}\,\Msun$, with disc compactness corresponding to redshift $z=0$--$10$, we find that more compact discs form more massive and denser clumps.
The maximum clump mass increases from $\sim 10^{5}$ to $\sim 10^{7}\,\Msun$, and the fraction of clumps exceeding the surface density of $300\,\mathrm{\Msun\,pc^{-2}}$ rises from $0.25$ to $0.38$ between $z=0$ and $10$ in $10^{10}\,\Msun$ haloes.
Star formation becomes correspondingly bursty, and the global star formation efficiency after three disc rotations increases from $5 \times 10^{-3}$ to $2 \times 10^{-1}$ in $10^{10}\,\Msun$ haloes and from $2 \times 10^{-2}$ to $4 \times 10^{-1}$ in $10^{11}\,\Msun$ haloes as the redshift increases from $z=0$ to $10$.
The accompanying feedback disrupts the gas discs of the most compact systems, while prominent stellar spiral arms emerge.
Self-gravitationally bound star clusters form only in compact discs.
We find no bound star clusters at $z \leq 2$ in $10^{10}\,\Msun$ haloes, whereas the bound cluster mass fraction reaches $\approx 40$ per cent at $z \geq 8$ and $\approx 70$ per cent in the most compact $10^{11}\,\Msun$ halo.
These fractions are broadly consistent with those inferred for high-$z$ clumpy galaxies observed by the James Webb Space Telescope.

\end{abstract}

\begin{keywords}
hydrodynamics -- methods: numerical -- galaxies: star formation -- galaxies: ISM
\end{keywords}



\section{Introduction}
\label{intro}

How galaxies convert gas into stars on the scale of individual clouds remains one of the central unsolved problems in galaxy formation.
Stars form within giant molecular clouds (GMCs), which are clumpy and dense structures embedded in the interstellar medium (ISM), and the efficiency of this conversion is regulated by the interplay between self-gravity and stellar feedback.
Once massive stars are born, stellar feedback processes such as photoionisation, radiation pressure, and supernova (SN) explosions inject energy, momentum, and metals into the surrounding ISM, dispersing their parent GMCs \citep[e.g.][]{kruijssen_2019,chevance_2022,schinnerer_2024,harer_2025,wainer_2026}. 
The dispersed gas subsequently cools, re-condenses, and forms new GMCs, where the next generation of stars emerges \citep[e.g.][]{inutsuka_2015}.
This continuous gas--star cycle regulates the evolution of galaxies from the early Universe to the present day.
Understanding how gas within GMCs is converted into stars is therefore essential for revealing the physical mechanisms that drive galaxy formation and evolution.

Clumpy morphologies in high-redshift star-forming galaxies have been recognised frequently.
Deep imaging with the Hubble Space Telescope revealed that a substantial fraction of galaxies at the peak epoch of cosmic star formation, at redshift $z \sim 1$--$3$, are dominated by a small number of giant clumps with sizes of $\sim 1 \, {\rm kpc}$ and stellar masses of $10^{8}$--$10^{9}\,\Msun$, in marked contrast to the smooth spiral discs seen in the local Universe \citep[e.g.][]{foster-schreiber_2009, guo_2015}.
These clumps are commonly interpreted as arising from gravitational instability in gas-rich, turbulent discs, in which a Toomre-unstable disc fragments into massive bound structures \citep[e.g.][]{dekel_2009_b}.
Whether such clumps survive stellar feedback long enough to migrate inwards and contribute to bulge growth, or are instead rapidly disrupted, has remained a matter of debate \citep[e.g.][]{genel_2012, oklopcic_2017}.
Understanding the physical conditions under which clumps form and survive is therefore a long-standing question for understanding galaxy formation in the early Universe.

Recent high-resolution observations with the James Webb Space Telescope (JWST) have revealed the presence of extremely compact star clusters in galaxies in the high-$z$ Universe.
For example, \citet{adamo_2024} discovered five compact star clusters within a region of only $\sim70\,\pc$ in the Cosmic Gems arc, a strongly lensed galaxy at $z\approx10$.
These clusters have effective radii of only $\sim1\,\pc$ and stellar masses of a few $10^6\,\Msun$, implying stellar surface densities of $\gtrsim10^5\,\Msun\,\pc^{-2}$.
Such densities exceed those of typical young star clusters in nearby galaxies and the Milky Way by several orders of magnitude \citep[e.g.][]{portegies-zwart_2010,brown_2021}.
Similarly, \citet{fujimoto_2025} observed the Cosmic Grapes, a lensed rotating disc galaxy at $z\approx6$, and identified at least 15 compact star-forming regions with effective radii of $\approx10$--$60\,\pc$ and stellar masses of $\approx10^6$--$8\times10^7\,\Msun$.
These regions exhibit extremely high gas surface densities of $\gtrsim10^3$--$10^5\,\Msun\,\pc^{-2}$, potentially weakening stellar feedback and enabling highly efficient star formation \citep[e.g.][]{finkelstein_2022,harikane_2023}.
Despite these observational breakthroughs, the physical mechanisms responsible for the formation of such compact star clusters in high-$z$ galaxies remain poorly understood.

Recently, \citet{fukushima_2021} performed radiation hydrodynamics simulations of star-forming GMCs and demonstrated that the compactness of star clusters is strongly regulated by the initial gas surface density of their host clouds.
By systematically varying the initial surface densities of GMCs, they investigated how cloud compactness affects star formation efficiency and the properties of the resulting star clusters.
They found that when the initial gas surface density exceeds a critical threshold of $\sim300\,\Msun\,\pc^{-2}$, compact star clusters form with sufficiently deep gravitational potentials to retain surrounding gas, including ionised hot gas.
This allows star formation to continue efficiently, leading to high star formation efficiencies above $0.3$.
In contrast, when the initial surface density is below the threshold, photoionisation feedback efficiently disperses the host GMC, resulting in diffuse star clusters and star formation efficiencies below $0.2$.
These results suggest that the gas surface density of GMCs is a key parameter governing star cluster formation and may provide a physical explanation for the compact star clusters observed in high-$z$ galaxies \citep[][]{fujimoto_2025}.

Using the simulation results of \citet{fukushima_2021}, \citet{yajima_2025} developed a star cluster formation model for analytical models of galactic discs.
In this model, GMCs with surface densities below the critical threshold of $\Sigma_\mathrm{th}=300\,\Msun\,\pc^{-2}$ exhibit inefficient star formation, with star formation efficiencies remaining below $0.1$ and scaling as the square root of the gas surface density.
Once the surface density exceeds the threshold, however, the star formation efficiency increases rapidly, reaching $\sim0.3$ at $\Sigma \sim 10^3\,\Msun\,\pc^{-2}$.
By incorporating the migration and accretion of newly formed star clusters based on this model, \citet{yajima_2025} successfully reproduced the presence of over-massive black holes in compact high-$z$ disc galaxies, whose black hole masses significantly exceed those expected from the local black hole--galaxy mass relation \citep[][see also e.g. \citeauthor{kormendy_2013} \citeyear{kormendy_2013}]{maiolino_2024}.
The results suggest that compact star cluster formation may play an important role in driving rapid black hole growth in high-$z$ galaxies.
More broadly, the model provides a promising framework for understanding the connection between star formation, cluster formation, and black hole evolution in compact galactic discs.

Numerical simulations are also a powerful tool for investigating the formation of star clusters in galaxies \citep[e.g.][]{agertz_2013,ricotti_2016,kim_2020,abe_2021,yajima_2023}.
For example, using cosmological simulations of high-$z$ galaxies, \citet{sugimura_2024} found that far-ultraviolet (FUV) radiation from Population\,II stars dissociates molecular hydrogen and suppresses gas cooling, promoting the formation of massive gas clouds.
Once these clouds become sufficiently massive to undergo gravitational collapse, they form compact and massive star clusters, triggering bursty star formation activity.
This mechanism may help explain the unexpectedly large population of luminous galaxies observed in the high-$z$ Universe \citep[e.g.][]{harikane_2023,harikane_2024,robertson_2023,hainline_2024,napolitano_2025}.
However, their simulations focused on a limited halo population, reaching halo masses of only $\sim10^8\,\Msun$ at $z\sim10$.
The formation of compact star clusters is likely to depend not only on local cloud-scale physics, but also on global galactic properties such as halo mass and $z$, because these parameters strongly influence the compactness and gas density of galactic discs \citep[][]{mo_1998}.
A systematic investigation across a wider range of galactic environments is therefore required to understand when and where compact star clusters preferentially form.

In this paper, we investigate how the compactness of galactic discs regulates star formation activity and star cluster formation.
To this end, we develop a new star formation model for galaxy simulations, motivated by the GMC-scale radiation hydrodynamics simulations of \citet{fukushima_2021} and the analytical framework of \citet{yajima_2025}.
Then, we apply it to isolated disc galaxies spanning halo masses of $M_{\rm h} = 10^{9}$, $10^{10}$, and $10^{11}\,\Msun$ and redshifts of $z = 0$--$10$.

In cosmological simulations, galaxies of the same halo mass at the same redshift can exhibit markedly different star formation histories, depending on their large-scale environment, merger history, and the timing and geometry of cold gas accretion.
While these processes are important in shaping real galaxies, they also introduce a large scatter that makes it difficult to isolate the effect of disc compactness itself.
By constructing idealised discs in which the halo mass and the redshift-dependent disc scale radius are specified as free parameters, we can vary compactness in a controlled manner and attribute the resulting differences in cloud properties, star formation efficiency, and bound cluster formation directly to compactness.
This controlled setup also allows us to reach the mass and spatial resolution required to identify individual gas clumps on-the-fly, which would be expensive across the same range of halo masses in a cosmological volume. 

This paper is organised as follows.
In Section~\ref{methods}, we describe the numerical methods, galaxy models, and the newly developed star formation model adopted in our simulations.
Section~\ref{results} presents the main results of this study, while Section~\ref{discussion} discusses their implications.
Finally, we summarise our conclusions in Section~\ref{conclusion}.
\section{Methods}
\label{methods}

\subsection{Numerical simulations}
\label{methods:sim}

We perform isolated disc galaxy simulations using the GADGET-3 code \citep[][]{springel_2005_gadget2}, which solves self-gravity and hydrodynamics with the smoothed particle hydrodynamics (SPH) method.
Our version of the code inherits the subgrid physics developed for a series of previous projects.
Much of the subgrid physics originates from the OWLS project \citep{schaye_2010} and the FiBY project \citep{johnson_2013,paardekooper_2015}, and the star formation and supernova (SN) feedback models were subsequently updated following the EAGLE project \citep{schaye_2015_eagle}.
These developments were incorporated into the version of the code used for the FOREVER22 project \citep{yajima_2022}, which investigates protocluster regions through cosmological simulations, and which forms the basis of the present work.
Our simulations therefore include a variety of subgrid physical processes relevant to galaxy evolution, such as radiative cooling, star formation, and stellar feedback.
The details of the star formation model are described in Section~\ref{methods:sf}.
The gravitational softening lengths for DM, star, and gas particles are set to $50$, $10$, and $1\,\pc$, respectively.

Gas cools radiatively through molecular hydrogen line cooling, Compton cooling, and metal-line cooling \citep[][]{wiersma_2009}.
The metal-line cooling rates are calculated using pre-computed tables generated with the CLOUDY code \citep[][]{ferland_2000}.
We also include heating by the ultraviolet background (UVB) radiation field that permeates the ISM \citep[e.g.][]{haardt_1996,faucher-gigure_2009}.
The UVB is assumed to penetrate gas with hydrogen number densities of $n_\mathrm{H}<0.01\,\cc$ \citep[][]{nagamine_2010,yajima_2012}.
To prevent artificial fragmentation in dense gas regions, we adopt an effective equation of state for gas particles with $n_\mathrm{H}>100\,\cc$, using an effective adiabatic index of $\gamma_\mathrm{eff}=4/3$ and a temperature floor of $T_\mathrm{f}=100\,\K$ \citep[][]{schaye_2015_eagle}.
This treatment stabilises unresolved dense gas against artificial collapse while preserving the large-scale structure of the ISM.

When a star particle is younger than $10\,\myr$, it radiatively heats surrounding gas particles through photoionisation feedback.
Ionised gas particles are identified by estimating the balance between the ionising photon production rate from a young star particle and the recombination rate of the surrounding gas.
Gas particles are ionised sequentially from neighbouring particles in order of increasing distance from the radiation source, and the temperature of the resulting {\HII} region is set to $3 \times 10^4~\rm K$.
In addition to photoionisation heating, we include the effect of radiation pressure on dust grains.
UV radiation emitted by young star particles is partially absorbed by dust, injecting momentum into the surrounding gas and enhancing stellar feedback.
We assume that dust consists of silicate grains with a typical size of $\sim0.1\,\micron$ and a dust-to-gas mass ratio of $\sim0.01$ \citep[][]{yajima_2017b}.
Stars more massive than $6\,\Msun$ are assumed to explode as SNe once the corresponding star particle reaches an age of $10\,\myr$. 
SN feedback is implemented using the stochastic thermal feedback model of \citet{dalla-vecchia_2012}, in which neighbouring gas particles are randomly selected and heated to $10^{7.5}\,\K$.
The heated gas subsequently expands because of its high thermal pressure, driving bubbles and galactic outflows that disperse dense gas and regulate subsequent star formation within the ISM.

\subsection{Galaxy models}
\label{methods:gal}

We generate the initial conditions for our galaxy simulations using the DICE code \citep[see also \citeauthor{perret_2014} \citeyear{perret_2014}]{perret_2016_dice}, which determines the initial particle distributions using a Metropolis--Hastings Markov Chain Monte Carlo algorithm \citep[][]{metropolis_1953}.
In our initial conditions, an isolated gaseous disc is embedded within a spherical DM halo.
To isolate the impact of gaseous disc compactness on star formation activity, we neglect an initial stellar disc component for simplicity.
Although we perform non-cosmological simulations, the haloes are assumed to exist at $z=0$, $2$, $4$, $6$, $8$, and $10$.
The initial metallicity of gas particles is set to $Z=0.1Z_{\sun}$, where $Z_{\sun}=0.013$ is the solar metallicity \citep[][]{asplund_2009}.
The physical properties of the initial conditions adopted in our simulations are summarised in Table~\ref{tab:ics}.

The halo masses adopted in this work are $M_\mathrm{h}=10^9$, $10^{10}$, and $10^{11}\,\Msun$, corresponding to the virial masses of the host haloes.
DM particles follow the Navarro--Frenk--White (NFW) profile \citep[][]{navarro_1997} with a concentration parameter of $c=10$.
The radial distribution of the halo is truncated at the virial radius, $R_\mathrm{vir}$.
The value of $R_\mathrm{vir}$ depends on the assumed $z$, calculated using the cosmological parameters $H_0=67.4\,\mathrm{km\,s^{-1}\,Mpc^{-1}}$, $\Omega_\mathrm{m}=0.315$, $\Omega_\mathrm{b}=0.049$, and $\Omega_\mathrm{\Lambda}=0.685$ \citep[][]{planck_collaboration_2020}.
For a fixed halo mass, higher-$z$ haloes have smaller virial radii and are therefore more compact.
Thus, the adopted $z$ effectively controls the compactness of the galaxy halo in our models.
The mass of an individual DM particle is fixed at $m_\mathrm{DM}=10^5\,\Msun$ in all simulations.

The mass of the gaseous disc is set to 1 per cent of that of the host halo.
Therefore, the initial gas disc masses are $M_\mathrm{d}=10^7$, $10^{8}$, and $10^{9}\,\Msun$ for $M_\mathrm{h}=10^9$, $10^{10}$, and $10^{11}\,\Msun$, respectively.
Although observations and galaxy formation models suggest a correlation between halo mass and the baryonic mass of central galaxies \citep[e.g.][]{papastergis_2012,moster_2013}, we fix the gas-to-halo mass ratio in order to isolate the impact of galactic compactness on star formation activity.

The initial gas particles follow an exponential radial profile of $\exp(-R_\mathrm{gal}/R_\mathrm{d})$ and a vertical profile of $\mathrm{sech}^2(z_\mathrm{gal}/z_\mathrm{d})$, 
where $R_\mathrm{gal}$ and $z_\mathrm{gal}$ are the radial and vertical coordinates, respectively, and $R_\mathrm{d}$ and $z_\mathrm{d}$ are the disc scale radius and scale height.
Assuming conservation of baryonic specific angular momentum, the disc scale radius is computed as
\begin{equation}
    R_\mathrm{d} = \lambda R_\mathrm{vir},
\end{equation}
where $\lambda=0.04$ is the halo spin parameter \citep[][]{mo_1998}.
The scale height is set to $z_\mathrm{d}=0.1R_\mathrm{d}$.
As a result, for a fixed halo mass, galaxies at higher $z$ have more compact gaseous discs.
The initial mass of an individual gas particle is $1000\,\Msun$, although gas particle masses can evolve during the simulations because of stellar feedback and mass return.

Each simulation is evolved until $t=3t_\mathrm{dyn}$, where $t_\mathrm{dyn}$ is defined as the orbital time at the disc scale radius, computed from the circumference divided by the rotation speed at that radius.
We output 101 simulation snapshots at equal time intervals between $t=0$ and $3t_\mathrm{dyn}$.

\begin{table}
 \caption{Names and physical properties of the initial conditions; the halo mass $M_\mathrm{h}$, redshift $z$, the virial radius of the halo $R_\mathrm{vir}$, the scale radius of the gas disc $R_\mathrm{d}$, and the dynamical time of the gas disc $t_\mathrm{dyn}$.}
 \label{tab:ics}
 \centering
 \begin{tabular*}{\columnwidth}{l@{\extracolsep{\fill}}cccccc}
  \hline
  Name & $M_\mathrm{h}\,[\Msun]$ & $z$ & $R_\mathrm{vir}\,[\mathrm{kpc}]$ & $R_\mathrm{d}\,[\mathrm{kpc}]$ & $t_\mathrm{dyn}\,[\mathrm{Myr}]$\\
  \hline
  Mh9z0 & $10^9$ & $0$ & $21.2$ & $0.847$ & $377$\\[2pt]
  Mh9z2 & $10^9$ & $2$ & $10.1$ & $0.404$ & $116$\\[2pt]
  Mh9z4 & $10^9$ & $4$ & $6.18$ & $0.247$ & $56.2$\\[2pt]
  Mh9z6 & $10^9$ & $6$ & $4.43$ & $0.177$ & $33.1$\\[2pt]
  Mh9z8 & $10^9$ & $8$ & $3.45$ & $0.138$ & $22.8$\\[2pt]
  Mh9z10 & $10^9$ & $10$ & $2.83$ & $0.113$ & $16.8$\\[2pt]
  Mh10z0 & $10^{10}$ & $0$ & $45.6$ & $1.83$ & $387$\\[2pt]
  Mh10z2 & $10^{10}$ & $2$ & $21.8$ & $0.870$ & $127$\\[2pt]
  Mh10z4 & $10^{10}$ & $4$ & $13.3$ & $0.533$ & $60.4$\\[2pt]
  Mh10z6 & $10^{10}$ & $6$ & $9.55$ & $0.382$ & $36.5$\\[2pt]
  Mh10z8 & $10^{10}$ & $8$ & $7.44$ & $0.297$ & $25.0$\\[2pt]
  Mh10z10 & $10^{10}$ & $10$ & $6.09$ & $0.243$ & $18.5$\\[2pt]
  Mh11z0 & $10^{11}$ & $0$ & $98.3$ & $3.93$ & $382$\\[2pt]
  Mh11z2 & $10^{11}$ & $2$ & $46.9$ & $1.88$ & $125$\\[2pt]
  Mh11z4 & $10^{11}$ & $4$ & $28.7$ & $1.15$ & $60.4$\\[2pt]
  Mh11z6 & $10^{11}$ & $6$ & $20.6$ & $0.823$ & $36.4$\\[2pt]
  Mh11z8 & $10^{11}$ & $8$ & $16.0$ & $0.641$ & $25.0$\\[2pt]
  Mh11z10 & $10^{11}$ & $10$ & $13.1$ & $0.525$ & $18.5$\\[2pt]
  \hline
 \end{tabular*}
\end{table}

\subsection{Star formation model of gas clumps}
\label{methods:sf}

To incorporate a star formation model based on the physical properties of clouds into galaxy simulations, it is necessary to identify GMCs on-the-fly during the simulations.
In this section, we describe the method used to extract GMCs in the simulations and the subgrid star formation model based on the compactness and clumpiness of the identified clouds.

\subsubsection{Extracting GMCs on-the-fly}
\label{methods:sf:gmcs}

We implement an on-the-fly GMC identification algorithm for galaxy simulations following \citet{horie_2024}.
This algorithm employs the Friends-of-Friends (FoF) method to group neighbouring dense gas particles into GMC candidates.
We adopt a threshold density of $n_\mathrm{H}=100\,\cc$ and a linking length equal to the gas smoothing length, $h_\mathrm{sml}$.
For gas particles with $n_\mathrm{H}=100\,\cc$, the typical smoothing length is $h_\mathrm{sml}\approx10\,\pc$ in our simulation setup.
A group containing at least $50$ dense gas particles is identified as a GMC.
Given this particle number threshold and the adopted gas mass resolution, the minimum resolved GMC mass is $\sim5\times10^4\,\Msun$.

The on-the-fly GMC identification is performed every $\Delta T_\mathrm{c}$, at which star formation is also calculated.
In this study, we adopt $\Delta T_\mathrm{c}=1\,\myr$.
Because this timescale is sufficiently shorter than the dynamical time of the simulated galaxies, $t_\mathrm{dyn}$, the simulations can self-consistently capture the interplay between GMC evolution, star formation, and subsequent stellar feedback.

We next define the global properties of the identified GMCs.
The GMC mass, $M_\mathrm{c}$, is defined as the total mass of dense gas particles belonging to the cloud.
The GMC radius, $R_\mathrm{c}$, is calculated from the total volume occupied by dense gas particles in the GMC, $V_\mathrm{c}$, as
\begin{equation}
    R_\mathrm{c} = \left( \frac{3V_\mathrm{c}}{4\pi} \right)^{1/3}.
\end{equation}
Using these quantities, we define the GMC surface density, $\Sigma_\mathrm{c}$, as
\begin{equation}
    \Sigma_\mathrm{c} = \frac{M_\mathrm{c}}{\pi R_\mathrm{c}^2}.
\end{equation}
The GMC surface density is a key parameter in our model and is used to estimate the star formation efficiency within each cloud.
The bulk velocity of a GMC, $\bar{\bm{\varv}}_\mathrm{c}$, is defined as
\begin{equation}
    \bar{\bm{\varv}}_\mathrm{c} = \langle \bm{\varv} \rangle_M,
\end{equation}
where $\langle \cdot \rangle_M$ denotes the mass-weighted average of a given quantity and $\bm{\varv}$ is the velocity of the member gas particles.
The one-dimensional velocity dispersion of a GMC is then given by
\begin{equation}
    \sigma_\mathrm{c}^2 = \frac{1}{3} \left\langle \left(\bm{\varv} - \bar{\bm{\varv}}_\mathrm{c}\right)^2 \right\rangle_M.
\end{equation}
In addition, the sound speed of a GMC is computed as
\begin{equation}
    c_{\mathrm{s,c}} = \sqrt{\langle c_\mathrm{s}^2 \rangle_M},
\end{equation}
where $c_\mathrm{s}$ is the sound speed of each gas particle.
We further define the virial parameter of a GMC as
\begin{equation}
    \alpha_\mathrm{vir} = \frac{5\left(\sigma_\mathrm{c}^2 + c_{\mathrm{s,c}}^2\right)R_{\mathrm{c}}}{G M_{\mathrm{c}}}, \label{eq:virialCloud}
\end{equation}
where $G$ is the gravitational constant \citep[][]{bertoldi_1992}.
A GMC with $\alpha_\mathrm{vir}<2$ is considered gravitationally bound.

\subsubsection{Conversion of gas to stars}
\label{methods:sf:gas-star}

Once GMCs are identified in our simulations every $\Delta T_\mathrm{c}=1\,\myr$, we compute the star formation efficiency, $\epsilon_\mathrm{SF}$, following \citet{yajima_2025} for self-gravitating GMCs ($\alpha_\mathrm{vir}<2$).
The functional form of $\epsilon_\mathrm{SF}$ is given by
\begin{equation}
\epsilon_{\mathrm{SF}} =
    \begin{cases}
        2.5 \times 10^{-2}\,
        \epsilon_{\mathrm{ff},0.02}^{4/5}\,
        T_{\mathrm{H}_\mathrm{II},4}^{-14/25}\,
        \Sigma_{\mathrm{c},100}^{1/2}
        & \text{for } \Sigma_{\mathrm{c}} < 300\,\Msun\,\mathrm{pc}^{-2} \\[6pt]
        0.5\left(
        \sqrt{a^2 \Sigma_{\mathrm{c},100}^{-2} + 4}
        - a\,\Sigma_{\mathrm{c},100}^{-1}
        \right)
    & \text{for } \Sigma_{\mathrm{c}} \ge 300\,\Msun\,\mathrm{pc}^{-2}
    \end{cases}
\hspace{2pt},
\label{eq:sfe}
\end{equation}
where $a=38.1$, $\epsilon_{\mathrm{ff},0.02}=\epsilon_{\mathrm{ff}}/0.02$, $T_{\mathrm{H}_\mathrm{II},4}=T_{\mathrm{H}_\mathrm{II}}/10^4\,\K$, and $\Sigma_{\mathrm{c},100}=\Sigma_{\mathrm{c}}/100\,\Msun\,\pc^{-2}$.
In this work, the star formation efficiency per free-fall time is fixed to $\epsilon_{\mathrm{ff}}=0.02$, and the \HII\, region temperature is assumed to be $T_{\mathrm{H}_\mathrm{II}}=10^4\,\K$, since stars are inefficiently formed in gas ionised by stellar radiation when $\Sigma_{\mathrm{c}} < 300\,\Msun\,\mathrm{pc}^{-2}$ \citep[][]{fukushima_2021}.
With these assumptions, Eq.~\ref{eq:sfe} depends only on the GMC surface density, $\Sigma_\mathrm{c}$.

This prescription reproduces the increase in star formation efficiency with increasing cloud surface density found by \citet{fukushima_2021}.
For $\Sigma_{\mathrm{c}} < 300\,\Msun\,\mathrm{pc}^{-2}$, $\epsilon_\mathrm{SF}$ remains at the level of a few per cent or less, reflecting efficient disruption of GMCs by photoionisation feedback.
In contrast, for $\Sigma_{\mathrm{c}} \ge 300\,\Msun\,\mathrm{pc}^{-2}$, $\epsilon_\mathrm{SF}$ increases to values of order a few tens of per cent, corresponding to continued star formation in compact, gravitationally bound clouds where feedback is less effective at dispersing the gas.
This behaviour is motivated by the picture that star formation in GMCs proceeds until radiation pressure becomes comparable to the self-gravity of the forming star cluster.

Using this star formation efficiency, $\epsilon_\mathrm{SF}$, we derive the stellar mass formed in a GMC with $\alpha_\mathrm{vir}<2$ as
\begin{equation}
    M_{\star,\mathrm{c}} = \epsilon_\mathrm{SF} M_\mathrm{c}.
\end{equation}
Star formation within a GMC then proceeds from its densest regions.
Specifically, we convert member gas particles into stars in order of decreasing density until the converted mass reaches $M_{\star,\mathrm{c}}$.
To implement this, for each member gas particle with density $\rho$, we compute the cumulative mass of all denser particles in the same GMC, $M(>\rho)$, and select those satisfying $M(>\rho) \le M_{\star,\mathrm{c}}$ as star-forming particles.
Each selected particle either spawns a new star particle or is converted into one, depending on its mass.
If its mass exceeds $1.5\,m_\mathrm{gas}$, where $m_\mathrm{gas}=1000\,\Msun$ is the initial mass of a single gas particle, it spawns a star particle of mass $m_\mathrm{gas}$ and retains the remaining mass as gas.
Otherwise, it is converted entirely into a star particle with the same mass as the original gas particle.
This procedure ensures that stars form from the densest gas in each cloud while conserving mass.

In our simulations, star particles are treated as simple stellar populations following a Chabrier initial mass function with a mass range of $0.1$-$100\,\Msun$ \citep[][]{chabrier_2003}.
While recent galaxy simulations typically adopt stochastic star formation models based on local gas properties \citep[e.g.][]{yajima_2022,hopkins_2023,horie_2024}, our approach is deterministic and directly based on the physical properties of the identified GMCs.

For comparison with our newly developed star formation model, we also perform simulations of the $10^{10}\,\Msun$ haloes using a conventional star formation model based on the Kennicutt--Schmidt (KS) law \citep[][]{schaye_2008}.
In this model, star formation rates (SFRs) are calculated for dense gas particles according to the observed KS relation \citep[][]{kennicutt_1998}, and gas particles are stochastically converted into star particles based on the resulting SFRs.
This approach has been widely adopted in previous galaxy simulations \citep[e.g.][]{schaye_2010,schaye_2015_eagle,yajima_2022}.
In the KS model, gas particles with $n_\mathrm{H}>100\,\cc$ are allowed to form stars.
Throughout this paper, we refer to simulations using our cloud-based star formation model as `{\it CSF}', while simulations adopting the KS-law-based model are denoted as `{\it KS}'.
Unless otherwise stated, we adopt the simulations with $M_\mathrm{h}=10^{10}\,\Msun$ and the {\it CSF} model as our fiducial runs.

\section{Results}
\label{results}

\subsection{Fiducial halo mass cases}
\label{results:fiducial}

We first present the results for our fiducial models, which adopt the initial conditions with $M_\mathrm{h}=10^{10}\,\Msun$ and the CSF model.

\subsubsection{Galactic morphologies and star formation}
\label{results:fiducial:morph_sf}

\begin{figure}
    \includegraphics[width=\columnwidth]{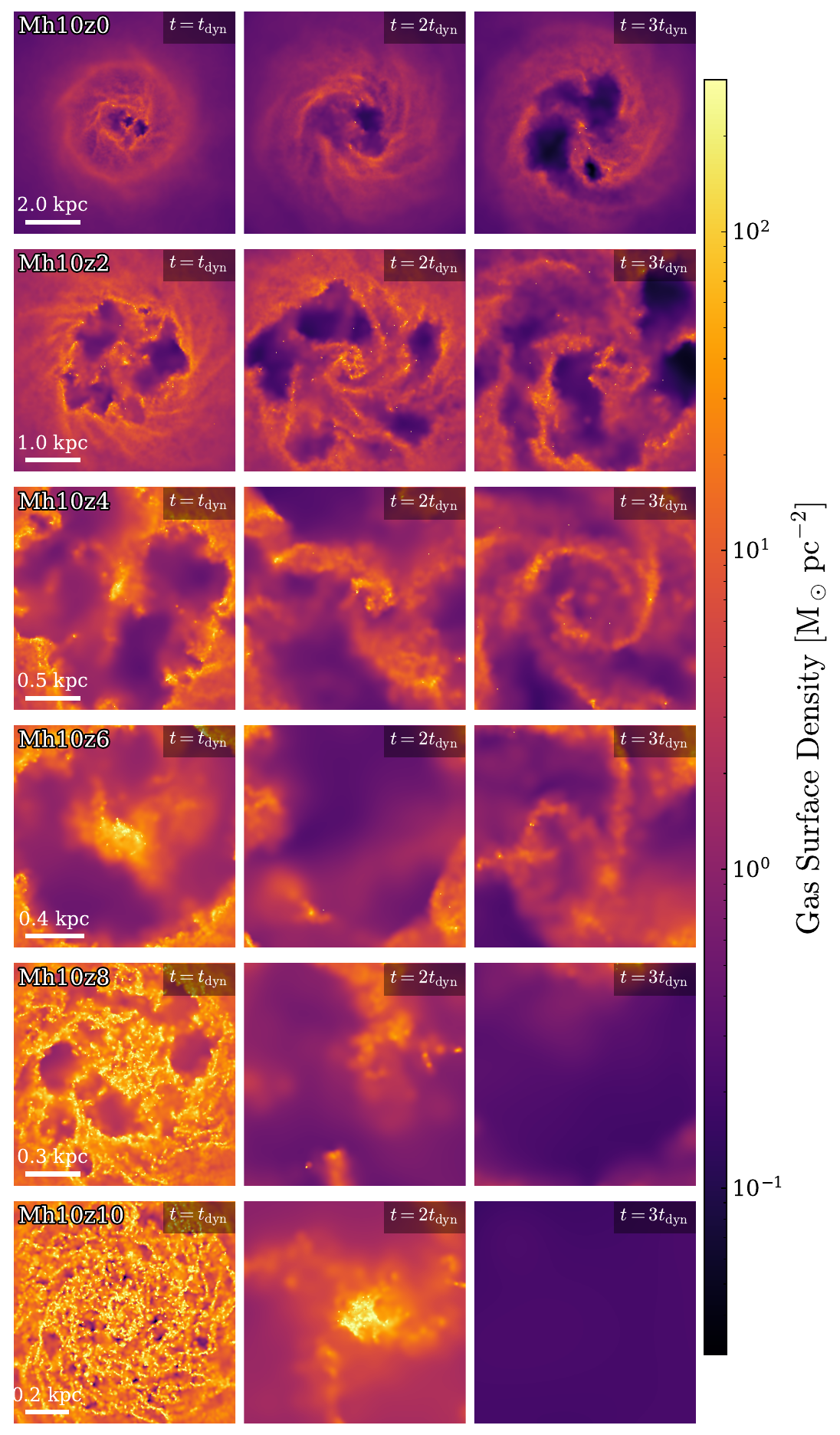}
    \caption{Time evolution of the gas surface density of the galaxies simulated using the initial conditions of $10^{10}\,\Msun$ haloes and the CSF model. Left, center, and right columns represent the face-on views at $t=t_\mathrm{dyn}, 2t_\mathrm{dyn}, 3t_\mathrm{dyn}$, respectively. The initial conditions used are shown in the upper left of the panels in the left column. Each panel is $\approx 4R_\mathrm{d}$ across, and the scale bars are also shown in the panels.}
    \label{fig:gassurf_Mh10_CSF}
\end{figure}

\begin{figure}
    \includegraphics[width=\columnwidth]{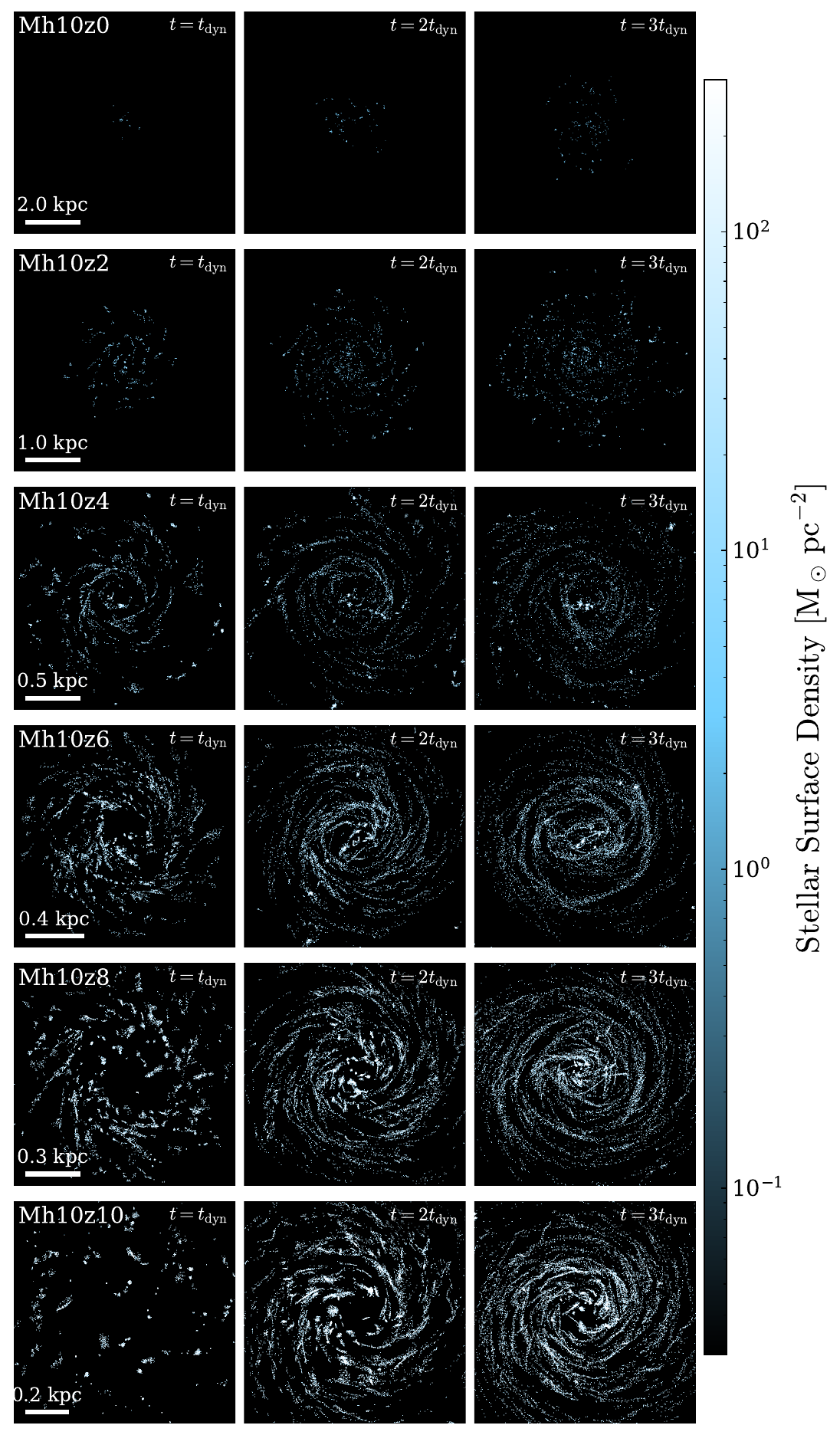}
    \caption{Time evolution of stellar surface density of the galaxies simulated using the initial conditions of $10^{10}\,\Msun$ haloes and the CSF model.}
    \label{fig:starsurf_Mh10_CSF}
\end{figure}

\begin{figure}
    \centering
    \includegraphics[width=\columnwidth]{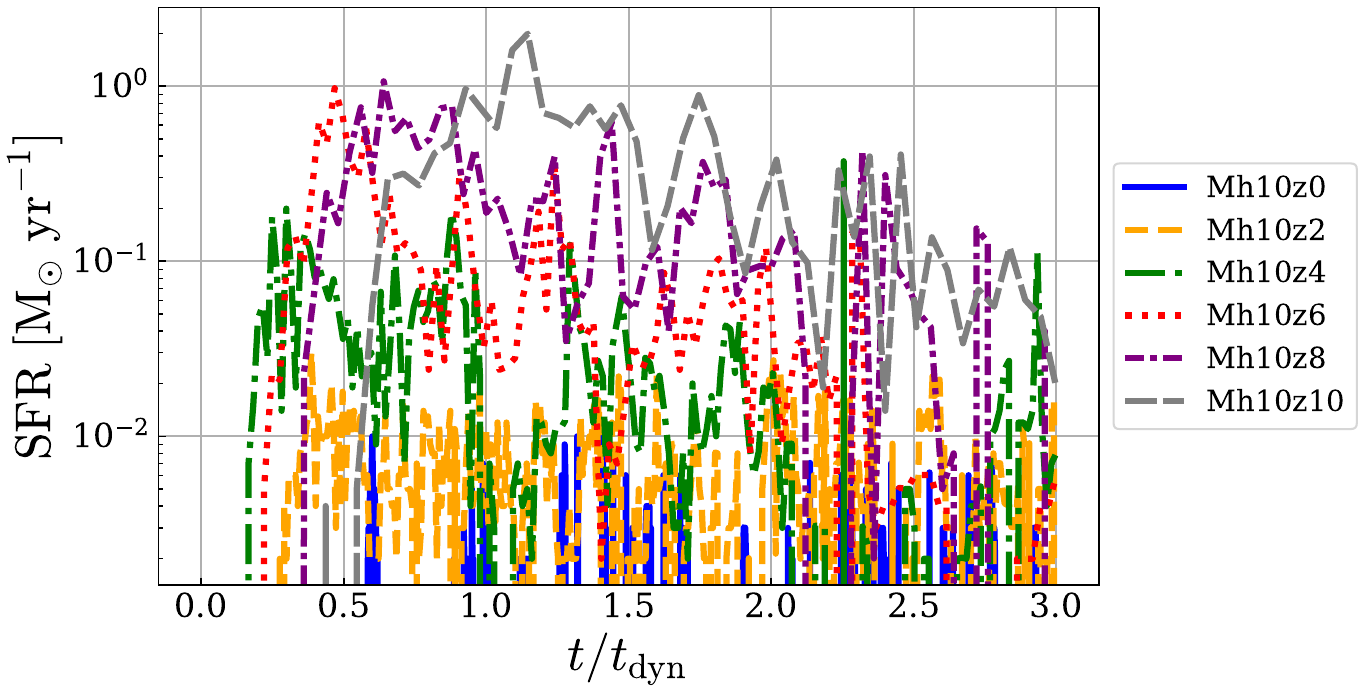}\\[2ex]
    \includegraphics[width=\columnwidth]{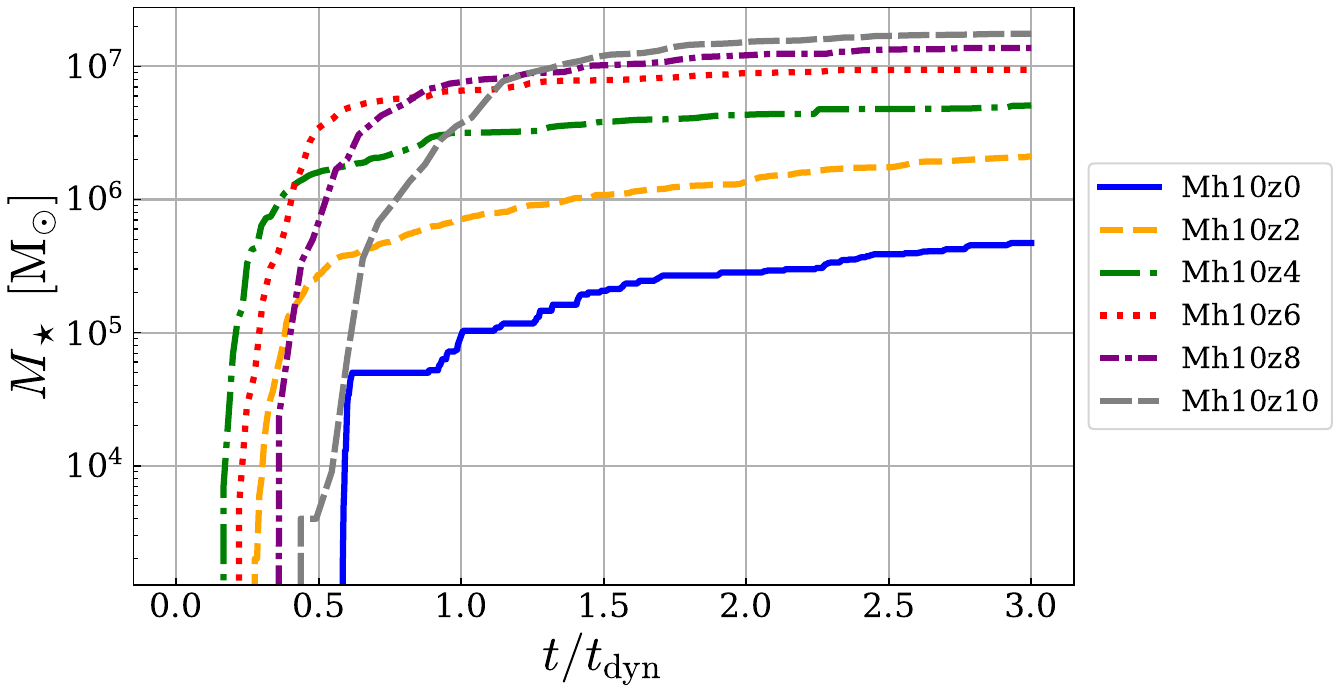}
    \caption{SFRs (top) and total stellar masses $M_\star$ (bottom) as a function of time for the simulations using the initial conditions of $10^{10}\,\Msun$ haloes and the CSF model. The lines represent the different initial conditions used as shown in the legends.}
    \label{fig:sfh_Mh10_CSF}
\end{figure}

Figure~\ref{fig:gassurf_Mh10_CSF} shows the evolution of the gas surface density in simulated galaxies with $10^{10}\,\Msun$ haloes using the CSF model.
For the galaxies at $z=0$ and $2$, the gaseous discs remain stable over three rotational periods, although kpc-scale hot bubbles driven by stellar feedback develop within the ISM.
In contrast, the galaxies at $z=4$ and $6$ experience substantial gas blowout caused by stellar feedback, after which part of the expelled gas falls back towards the galactic disc until $t=3t_\mathrm{dyn}$.
The galaxies at $z=8$ and $10$ show even more dramatic evolution.
Their gaseous structures cannot be maintained at $t\gtrsim2t_\mathrm{dyn}$, and little gas remains around the galactic centre by $t=3t_\mathrm{dyn}$.
These results indicate that stellar feedback becomes increasingly effective at disrupting gaseous discs in $10^{10}\,\Msun$ haloes towards higher $z$, particularly for $z\gtrsim4$.

Although gaseous structures become increasingly disrupted in higher-$z$ galaxies, the stellar component develops prominent spiral structures.
Figure~\ref{fig:starsurf_Mh10_CSF} shows face-on maps of the stellar surface density in the simulated galaxies.
While the galaxy at $z=0$ does not exhibit clear stellar spiral structures, such features become progressively more pronounced towards higher $z$.
Although the spiral patterns are relatively weak at $t=t_\mathrm{dyn}$, they gradually grow and elongate over time.
These results suggest that compact high-$z$ discs efficiently form dynamically coherent stellar structures despite strong stellar feedback acting on the gaseous component.
However, the gradual disappearance of compact stellar structures may partly reflect the effect of gravitational softening, which can artificially disrupt self-gravitating compact star clusters.

The impact of galactic compactness is also clearly reflected in the star formation activity.
The top panel of Fig.~\ref{fig:sfh_Mh10_CSF} shows the star formation rates (SFRs) over three rotational periods.
While the $z=0$ and $2$ galaxies exhibit intermittent star formation activity, the higher-$z$ galaxies show more continuous and intense star formation throughout the simulations.
In the $z=4$ case, the SFR remains roughly between $10^{-2}$ and $10^{-1}\,\sfrunit$.
In contrast, the galaxies at $z=6$, $8$, and $10$ reach peak SFRs of $\sim1\,\sfrunit$ until $t\approx t_\mathrm{dyn}$, after which the SFRs gradually decline.
Such bursty star formation drives strong stellar feedback, injecting large amounts of energy into the surrounding ISM and ultimately disrupting the gaseous structures, as shown in Fig.~\ref{fig:gassurf_Mh10_CSF}.
Such bursty star formation may also be relevant to the statistical properties of high-$z$ galaxies revealed by JWST.
The unexpectedly high abundance of UV-luminous galaxies at $z \gtrsim 10$ \citep[e.g.][]{harikane_2023} has been attributed in part to stochastic star formation, which scatters galaxies to brighter magnitudes and boosts the observed UV luminosity function.
Our model produces such burstiness self-consistently from cloud-scale physics, and thus provides a physically motivated origin for the scatter required by these interpretations.

The bottom panel of Fig.~\ref{fig:sfh_Mh10_CSF} shows the evolution of the total stellar mass formed during the simulations.
In all cases, the stellar mass growth largely saturates by $t\approx t_\mathrm{dyn}$.
Given the initial gas disc mass of $M_\mathrm{d}=10^8\,\Msun$, the galactic star formation efficiencies at $t=3t_\mathrm{dyn}$, defined as $\mathrm{SFE} = {M_\star(t=3t_\mathrm{dyn})}/{M_\mathrm{d}}$, are approximately $5\times10^{-3}$, $2\times10^{-2}$, $5\times10^{-2}$, $1\times10^{-1}$, $1\times10^{-1}$, and $2\times10^{-1}$ for $z=0$, $2$, $4$, $6$, $8$, and $10$, respectively.
These results indicate that galactic star formation becomes substantially more efficient in compact high-$z$ galaxies.

\subsubsection{Gas clumps}
\label{results:fiducial:clumps}

\begin{figure*}
  \centering

  \includegraphics[width=0.33\textwidth]{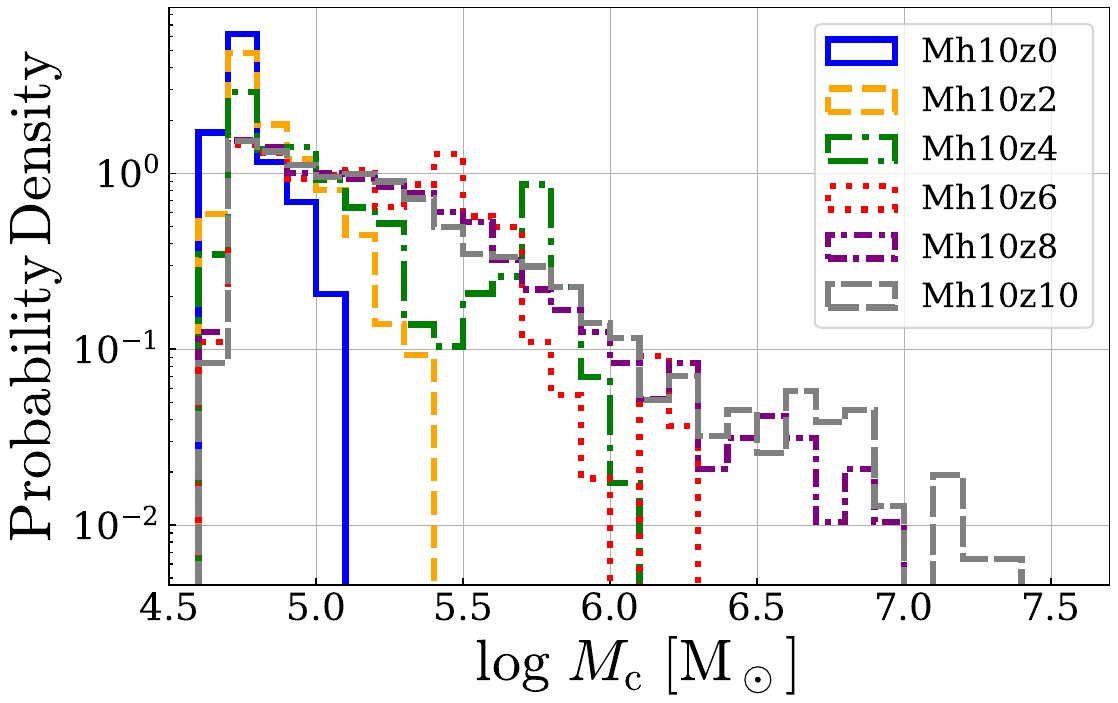}
  \includegraphics[width=0.33\textwidth]{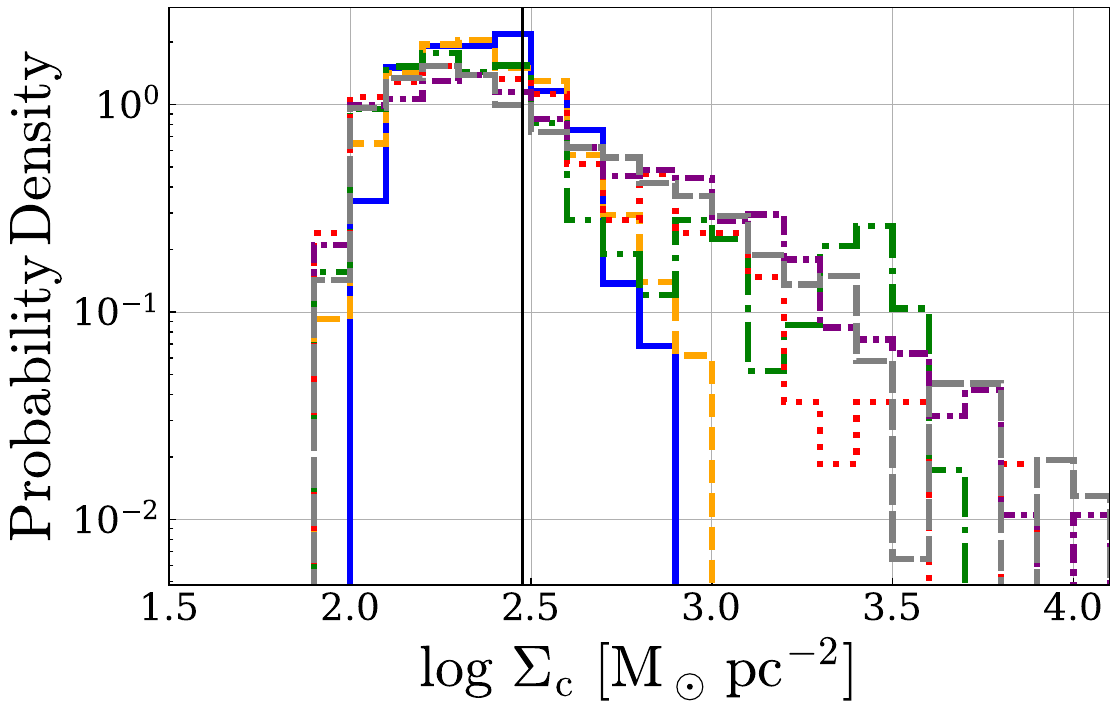}
  \includegraphics[width=0.33\textwidth]{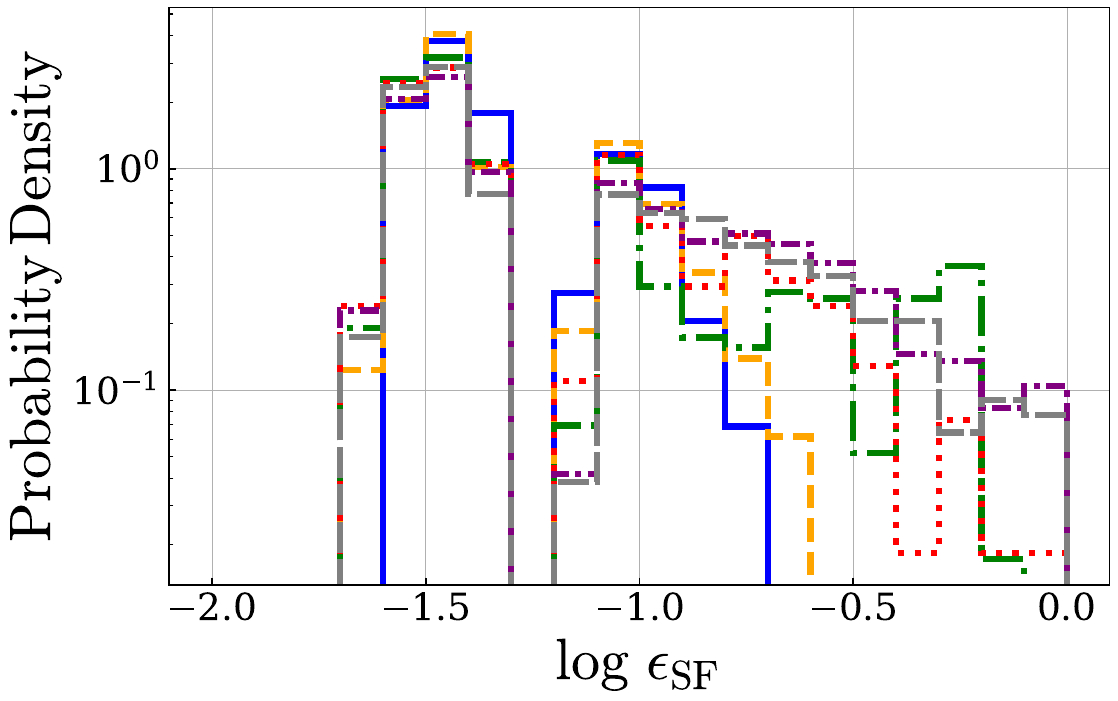}

  \caption{Probability distribution functions of the physical properties of gas clumps identified by the GMC identification algorithm described in Sec.~\ref{methods:sf:gmcs} for the simulations with $10^{10}\,\Msun$ haloes using the CSF model.
  The left, centre, and right panels show the distributions of the clump mass $M_\mathrm{c}$, surface density $\Sigma_\mathrm{c}$, and star formation efficiency $\epsilon_\mathrm{SF}$, respectively.
  The black vertical line in the central panel indicates the threshold surface density of $\Sigma_\mathrm{c,th}=300\,\Msun\,\mathrm{pc}^{-2}$, above which $\epsilon_\mathrm{SF}$ increases sharply (see Sec.~\ref{methods:sf:gas-star} for details).}
  \label{fig:gas_clump_phys_pdf}
\end{figure*}

The left panel of Fig.~\ref{fig:gas_clump_phys_pdf} shows the one-dimensional probability distribution functions (PDFs) of the masses, $M_\mathrm{c}$, of identified gas clumps.
Because of the particle-number threshold adopted in the GMC identification algorithm, gas clumps with $M_\mathrm{c}\lesssim5\times10^{4}\,\Msun$ cannot be identified.
We find that the mass distributions strongly depend on the compactness of the host galaxies.
In all simulations, the PDFs peak at $M_\mathrm{c}\approx10^{4.8}\,\Msun$ which is close to the minimum mass of identified GMCs, whereas the maximum clump mass increases significantly towards higher $z$.
The maximum masses reach $\sim10^{5.1}$, $10^{6.3}$, and $10^{7.4}\,\Msun$ for the $z=0$, $6$, and $10$ cases, respectively.
Although massive clumps become progressively rarer at higher masses, the overall mass distributions broaden substantially in more compact galaxies.
These results indicate that compact high-$z$ galaxies preferentially form more massive GMCs over a wider mass range.

The centre panel of Fig.~\ref{fig:gas_clump_phys_pdf} shows the PDFs of the gas surface density, $\Sigma_\mathrm{c}$, of the identified gas clumps.
This quantity is the key parameter that determines the star formation efficiency of each clump (see Sec.~\ref{methods:sf:gas-star}).
Although the PDFs in all simulations peak at $\sim10^{2.3}\,\Msun\,\mathrm{pc}^{-2}$, the distributions of $\Sigma_\mathrm{c}$ show a systematic dependence on galactic compactness.
In particular, the maximum surface densities increase significantly towards higher $z$, reaching $\sim10^{2.9}$, $10^{3.9}$, and $>10^{4.0}\,\Msun\,\pc^{-2}$ for the $z=0$, $6$, and $10$ simulations, respectively.
The fractions of gas clumps with surface densities above the threshold value of $\Sigma_\mathrm{c,th}=300\,\Msun\,\mathrm{pc}^{-2}$, above which the star formation efficiency exceeds $\sim10$ per cent, are $\sim0.25$, $0.34$, and $0.38$ for the $z=0$, $6$, and $10$ simulations, respectively.
These results indicate that gas clumps become systematically denser and more compact in higher-$z$ galaxies, leading to conditions favourable for efficient star formation.

As a result of the gas surface density distribution of the identified gas clumps, the star formation efficiency $\epsilon_\mathrm{SF}$ can be estimated via Eq.~\ref{eq:sfe}.
The right panel of Fig.~\ref{fig:gas_clump_phys_pdf} shows the PDFs of $\epsilon_\mathrm{SF}$ in our simulations.
Due to the sharp increase of $\epsilon_\mathrm{SF}$ at $\Sigma_\mathrm{c}=\Sigma_\mathrm{c,th}=300\,\Msun\,\mathrm{pc}^{-2}$, the distributions exhibit a deficit around $\epsilon_\mathrm{SF}\approx10^{-1.3}$.
Since Eq.~\ref{eq:sfe} is a monotonically increasing function of $\Sigma_\mathrm{c}$, the PDFs of $\epsilon_\mathrm{SF}$ directly reflect those of $\Sigma_\mathrm{c}$.
In all simulations, we find gas clumps with high star formation efficiencies of $\epsilon_\mathrm{SF}\gtrsim0.1$, originating from compact clumps with $\Sigma_\mathrm{c}\ge\Sigma_\mathrm{c,th}$.
In particular, the $z=6$, $8$, and $10$ simulations produce clumps with extremely high efficiencies of $\sim1$.
Such highly efficient star formation in dense clumps may be responsible for the bursty star formation activity at the galactic scale, especially in higher-$z$ galaxies, as discussed in Sec.~\ref{results:fiducial:morph_sf}.

\subsubsection{Bound star clusters}
\label{results:fiducial:clusters}

We here present the results for the formed star clusters, with particular focus on self-gravitationally bound systems.
However, due to the effect of gravitational softening, compact star clusters formed from highly efficient gas clumps may artificially disperse, even though such clusters are expected to remain compact \citep[][]{fukushima_2021}.
Therefore, we define self-gravitationally bound star clusters as those formed in gas clumps with a star formation efficiency of $\epsilon_\mathrm{SF}\ge0.3$, rather than using the ratio of kinetic to potential energy.
The value corresponds to the gas surface density of $\Sigma_\mathrm{c} \gtrsim 1.3\times10^3\,\Msun\,\pc^{-2}$.
This choice is motivated by the fact that star clusters formed in clouds with such high efficiencies are expected to remain self-gravitationally bound \citep[][]{baumgardt_2007}.

\begin{figure}
    \centering
    \includegraphics[width=\columnwidth]{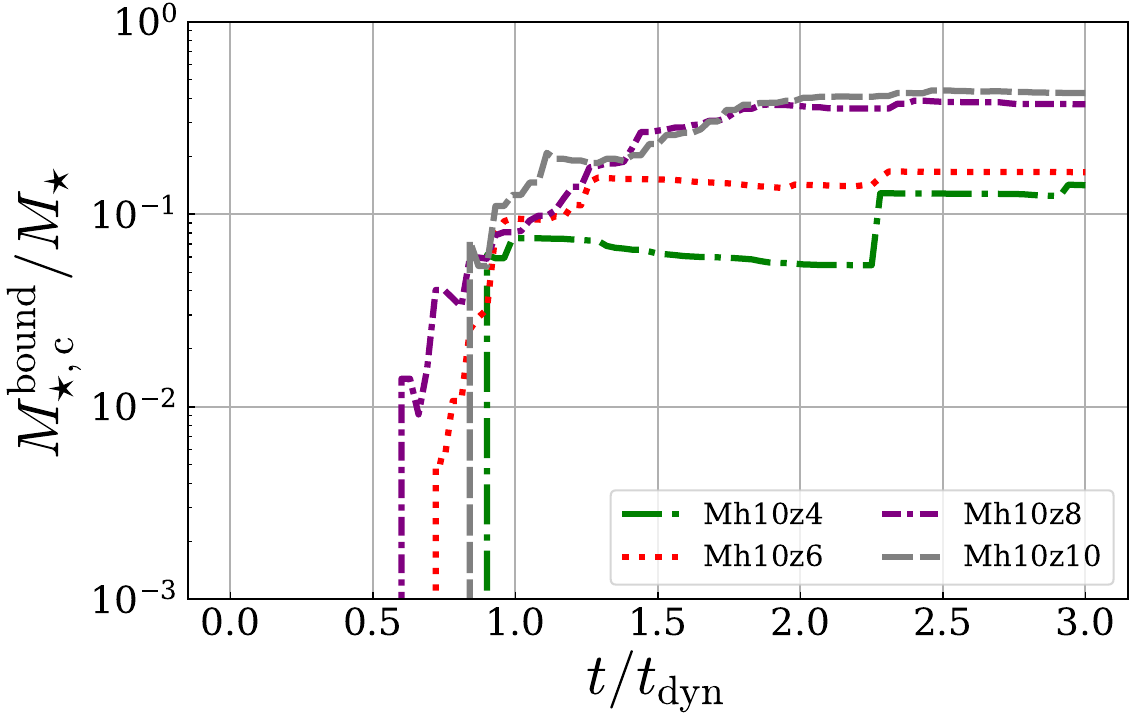}\\[2ex]
    \includegraphics[width=\columnwidth]{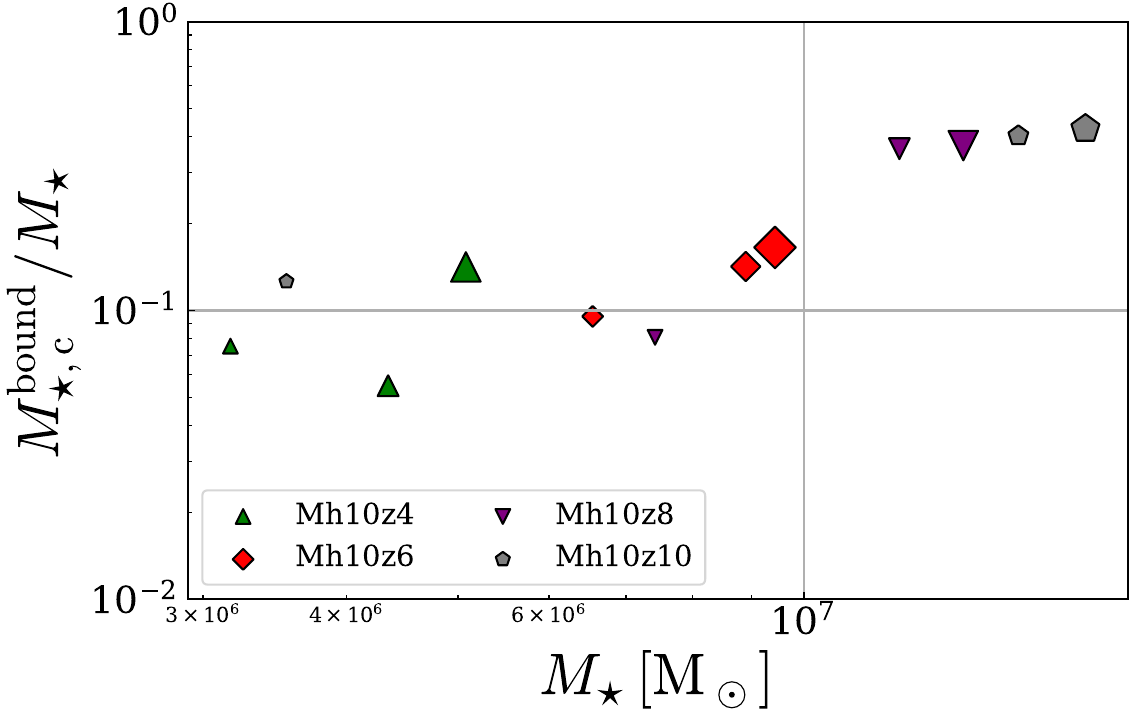}
    \caption{Top: Time evolution of the fraction of bound stellar cluster mass, $M_\mathrm{\star,c}^\mathrm{bound}$, relative to the total stellar mass, $M_\mathrm{\star}$, in the $M_\mathrm{h}=10^{10}\,\Msun$ simulations. The results for the $z=0$ and $2$ cases are not shown, as no self-gravitationally bound star clusters are formed in these runs. Bottom: The fraction $M_\mathrm{\star,c}^\mathrm{bound}/M_\mathrm{\star}$ as a function of the total stellar mass $M_\mathrm{\star}$. The markers indicate the results at $t=t_\mathrm{dyn}$, $2t_\mathrm{dyn}$, and $3t_\mathrm{dyn}$, with larger markers corresponding to later times.}
    \label{fig:bound_cluster_Mh10_CSF}
\end{figure}

The top panel of Fig.~\ref{fig:bound_cluster_Mh10_CSF} shows the time evolution of the fraction of the total mass in self-gravitationally bound star clusters, $M_\mathrm{\star,c}^\mathrm{bound}$, relative to the total stellar mass, $M_\mathrm{\star}$.
We find that this fraction strongly depends on galactic compactness.
In the $z=0$ and $2$ simulations, no self-gravitationally bound clusters form throughout the entire simulation time, indicating that star clusters in these systems originate from relatively diffuse gas clumps.
In contrast, the $z=4$, $6$, $8$, and $10$ simulations do form self-gravitationally bound star clusters.
At $t\approx t_\mathrm{dyn}$, the bound cluster mass fractions in these higher-$z$ simulations reach $\sim10\,\%$.
In the $z=4$ case, the fraction declines gradually between $t\approx t_\mathrm{dyn}$ and $\approx2.2t_\mathrm{dyn}$ and then increases sharply, driven by bursty star formation activity, as shown in the top panel of Fig.~\ref{fig:sfh_Mh10_CSF}
It appears that gas blown away by a previous episode of star formation and associated stellar feedback returns to the central region, where it forms a high-density gas clump that subsequently triggers this starburst.
For the $z=6$ simulation, the increase occurs mainly after $t=t_\mathrm{dyn}$, whereas in the $z=8$ and $10$ cases the fraction grows more gradually until $t\approx2t_\mathrm{dyn}$, eventually reaching values of $\approx40\,\%$.
These results demonstrate that both the formation efficiency and the time evolution of bound star clusters depend sensitively on galactic compactness.

The bottom panel of Fig.~\ref{fig:bound_cluster_Mh10_CSF} shows the relationship between the fraction $M_\mathrm{\star,c}^\mathrm{bound}/M_\mathrm{\star}$ and the total stellar mass $M_\mathrm{\star}$.
We find a weak positive correlation between the bound cluster mass fraction and the total stellar mass.
As time progresses, the total stellar mass increases and the bound fraction generally becomes higher, with the exception of the $z=4$ simulation.
In addition, more compact galaxies tend to exhibit both larger stellar masses and higher bound cluster fractions at $t=2t_\mathrm{dyn}$ and $3t_\mathrm{dyn}$.
These results suggest a systematic dependence of this relation on $z$, reflecting the role of galactic compactness in regulating bound star cluster formation.

\subsection{Dependence on halo masses}
\label{results:halo_mass}

We present the results of the simulations with $M_\mathrm{h}=10^{9}$ and $10^{11}\,\Msun$ haloes to examine the dependence on halo mass of galactic morphology, star formation activity, and the properties of star clusters.

\subsubsection{Galactic morphologies and star formation}
\label{results:halo_mass:morph_sf}

\begin{figure}
    \includegraphics[width=\columnwidth]{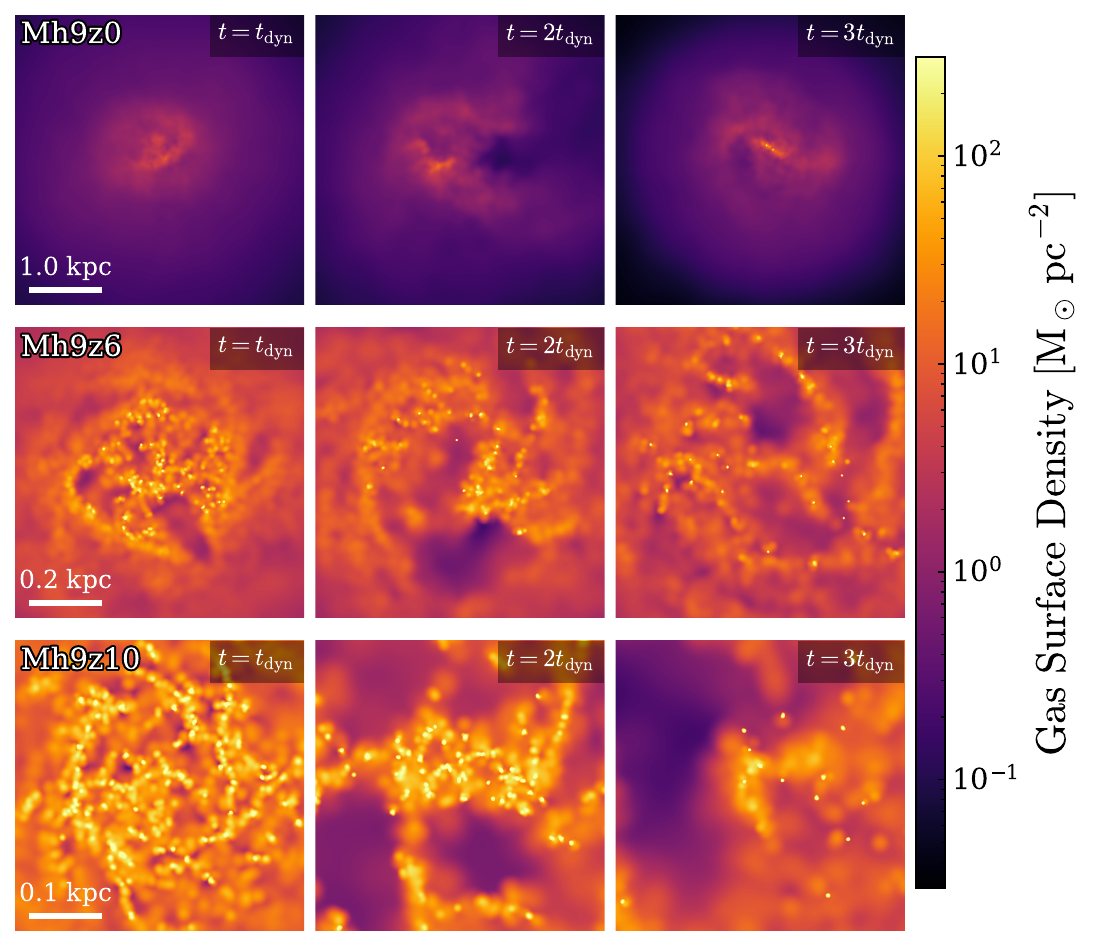}\\[2ex]
    \includegraphics[width=\columnwidth]{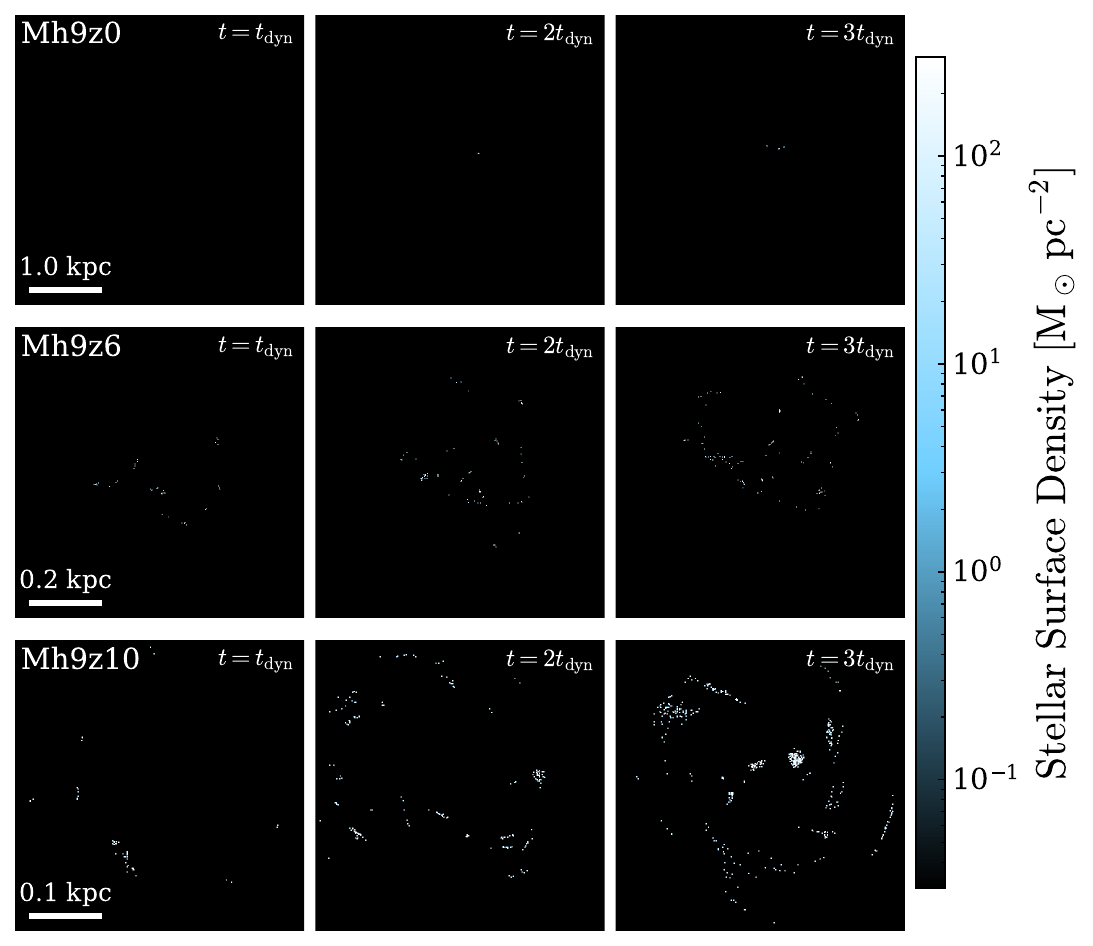}
    \caption{Time evolution of gas surface density (top) and stellar surface density (bottom) of the galaxies simulated using $10^{9}\,\Msun$ haloes. The results of the $z=0, 6$ and $10$ simulations are shown.}
    \label{fig:surfden_Mh9_CSF}
\end{figure}

We present the time evolution of the gas surface density and stellar surface density for galaxies simulated with $10^{9}\,\Msun$ haloes in the top and bottom panels of Fig.~\ref{fig:surfden_Mh9_CSF}.
We show three representative cases with $z=0$, $6$, and $10$ to illustrate the dependence of galactic morphology on halo mass.
In the $z=0$ simulation, the gaseous disc maintains its structure over three rotational periods without generating hot bubbles as large as those seen in the $10^{10}\,\Msun$ halo case, and stars are only weakly formed within the disc.
While the gas disc in the $z=6$ case for the $10^{10}\,\Msun$ halo is strongly disrupted (Fig.~\ref{fig:gassurf_Mh10_CSF}), the $z=6$ simulation with the $10^{9}\,\Msun$ halo also maintains a relatively stable disc structure over three rotations.
However, stellar spiral structures are not clearly observed, in contrast to the $10^{10}\,\Msun$ halo simulation.
In the $z=10$ simulation, gas remains within the disc region at $t=3t_\mathrm{dyn}$, whereas in the $10^{10}\,\Msun$ halo case the gas is almost completely blown out.
Stellar spiral structures are less distinct than in the $10^{10}\,\Msun$ halo simulation, although faint features can still be seen.
These results indicate that $10^{9}\,\Msun$ haloes exhibit systematically different morphological evolution compared to $10^{10}\,\Msun$ haloes.

\begin{figure}
    \centering
    \includegraphics[width=\columnwidth]{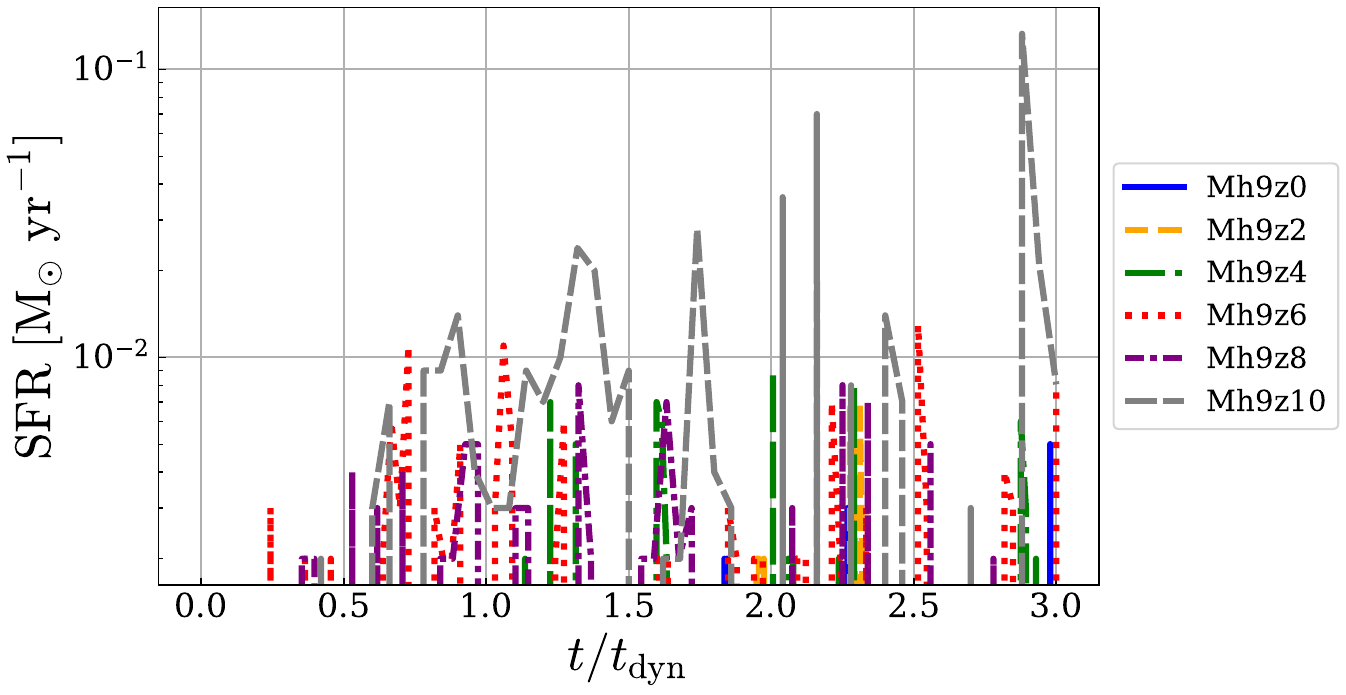}\\[2ex]
    \includegraphics[width=\columnwidth]{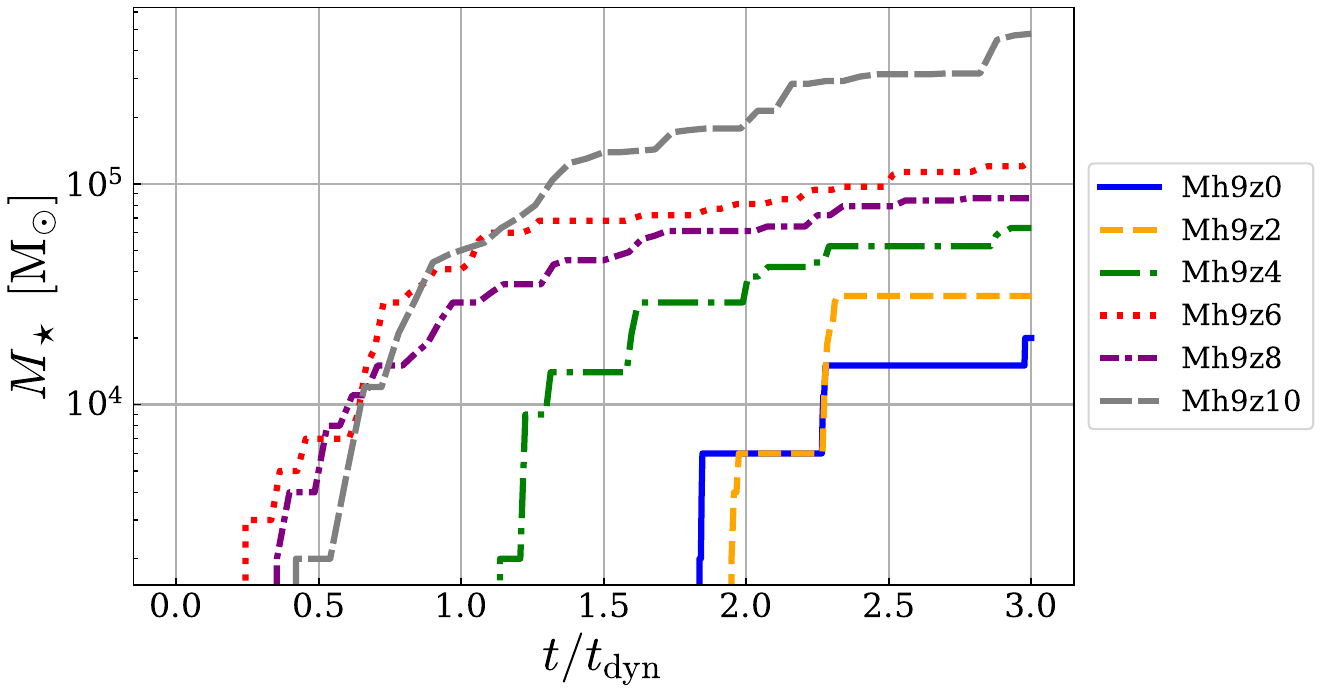}
    \caption{Same as Fig.~\ref{fig:sfh_Mh10_CSF}, but for the $10^{9}\,\Msun$ haloes.}
    \label{fig:sfh_Mh9_CSF}
\end{figure}

The difference between $10^{9}\,\Msun$ and $10^{10}\,\Msun$ haloes is also evident in the star formation activity.
Figure~\ref{fig:sfh_Mh9_CSF} shows the SFRs and total stellar masses as a function of time.
In the $10^{9}\,\Msun$ halo simulations, all runs exhibit intermittent star formation, whereas in the $10^{10}\,\Msun$ halo case, only the $z=0$ and $2$ runs show such intermittency.
The number and amplitude of star formation bursts are broadly correlated with galactic compactness.
For the $z=0$, $2$, $4$, $6$, and $8$ cases, the intermittent SFRs remain $\lesssim10^{-2}\,\sfrunit$, while the $z=10$ simulation reaches values above $10^{-2}\,\sfrunit$, peaking at $\sim10^{-1}\,\sfrunit$.
These differences in star formation activity indicate that feedback-driven regulation of the ISM depends sensitively on galactic compactness in $10^{9}\,\Msun$ haloes, leading to systematic variations in morphology.

The stellar mass growth in all simulations proceeds in a stepwise manner, reflecting the bursty nature of star formation, and generally saturates by $t\approx3t_\mathrm{dyn}$.
At $t=3t_\mathrm{dyn}$, the total stellar mass $M_\star$ roughly correlates with galactic compactness, although the value for the $z=6$ run is slightly higher than that for $z=8$.
The corresponding galactic star formation efficiencies at $t=3t_\mathrm{dyn}$ are $\mathrm{SFE}\approx2\times10^{-3}$, $3\times10^{-3}$, $7\times10^{-3}$, $9\times10^{-3}$, $1\times10^{-2}$, and $5\times10^{-2}$ for $z=0$, $2$, $4$, $6$, $8$, and $10$, respectively.
This result suggests that disc galaxies in $10^{9}\,\Msun$ haloes form stars significantly less efficiently than those in $10^{10}\,\Msun$ haloes, by factors of several to an order of magnitude.

\begin{figure}
    \includegraphics[width=\columnwidth]{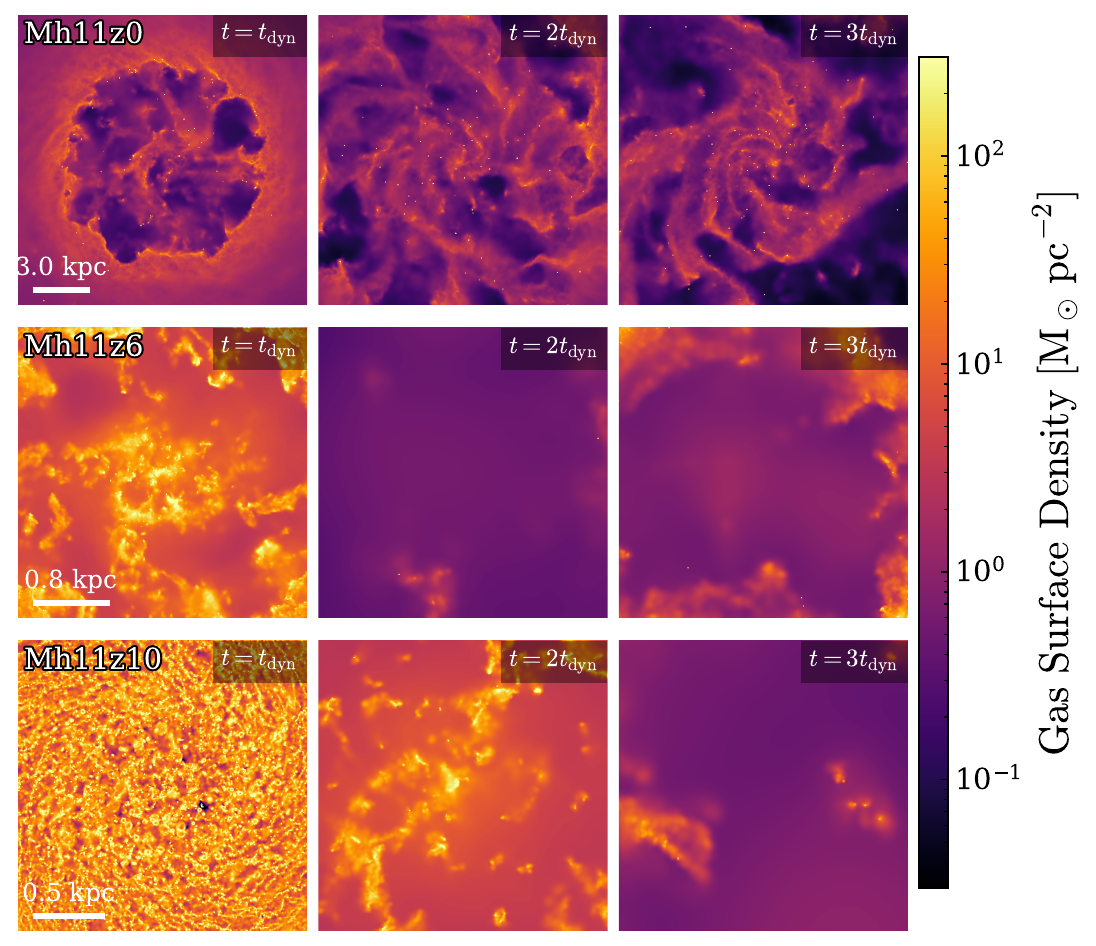}\\[2ex]
    \includegraphics[width=\columnwidth]{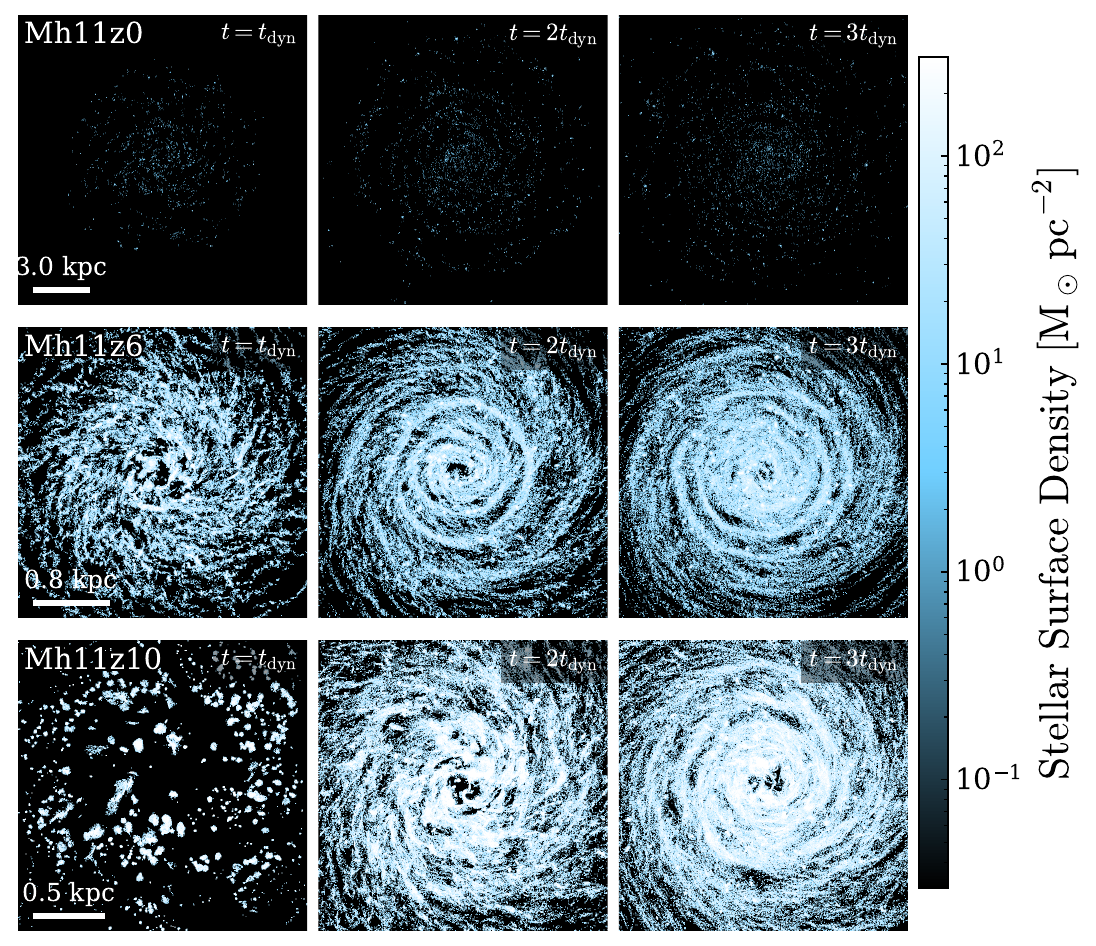}
    \caption{Same as Fig.~\ref{fig:surfden_Mh9_CSF}, but for the $10^{11}\,\Msun$ halo simulations.}
    \label{fig:surfden_Mh11_CSF}
\end{figure}

We also present the results of the $10^{11}\,\Msun$ halo simulations.
Figure~\ref{fig:surfden_Mh11_CSF} shows the gas surface density and stellar surface density of galaxies embedded in $10^{11}\,\Msun$ haloes.
In the $z=0$ simulation, although the gas disc is strongly affected by stellar feedback, it remains gravitationally bound near the galactic centre over three rotations, and clear spiral structures are present.
In the $z=6$ case, the disc is disrupted by stellar feedback at $t=t_\mathrm{dyn}$.
Although gas gradually returns toward the galactic centre by $t=3t_\mathrm{dyn}$, a coherent disc structure is not re-established.
As in the $10^{10}\,\Msun$ halo case, the $z=10$ disc is completely destroyed, and the gas does not significantly reaccrete.
Overall, the trends in the $10^{11}\,\Msun$ halo simulations are broadly similar to those in the $10^{10}\,\Msun$ haloes, indicating that feedback-driven disc disruption becomes increasingly severe at higher $z$ regardless of halo mass, within the range explored here.

Their stellar morphological evolution is similar to that in the $10^{10}\,\Msun$ halo cases, although with some notable differences.
While the $z=0$ simulation does not show clear spiral arms, the other runs produce spiral structures with higher stellar surface densities than those seen in the $10^{10}\,\Msun$ halo cases.
These spiral features are relatively weak at $t=t_\mathrm{dyn}$ but gradually develop into rotating stellar discs by $t=3t_\mathrm{dyn}$.
These results suggest that stellar discs can form rapidly from $10^{9}\,\Msun$ gas discs embedded in $10^{11}\,\Msun$ haloes, on timescales of $\sim100\,\myr$ in the very high-$z$ ($z\ge6$) Universe.

\begin{figure}
    \centering
    \includegraphics[width=\columnwidth]{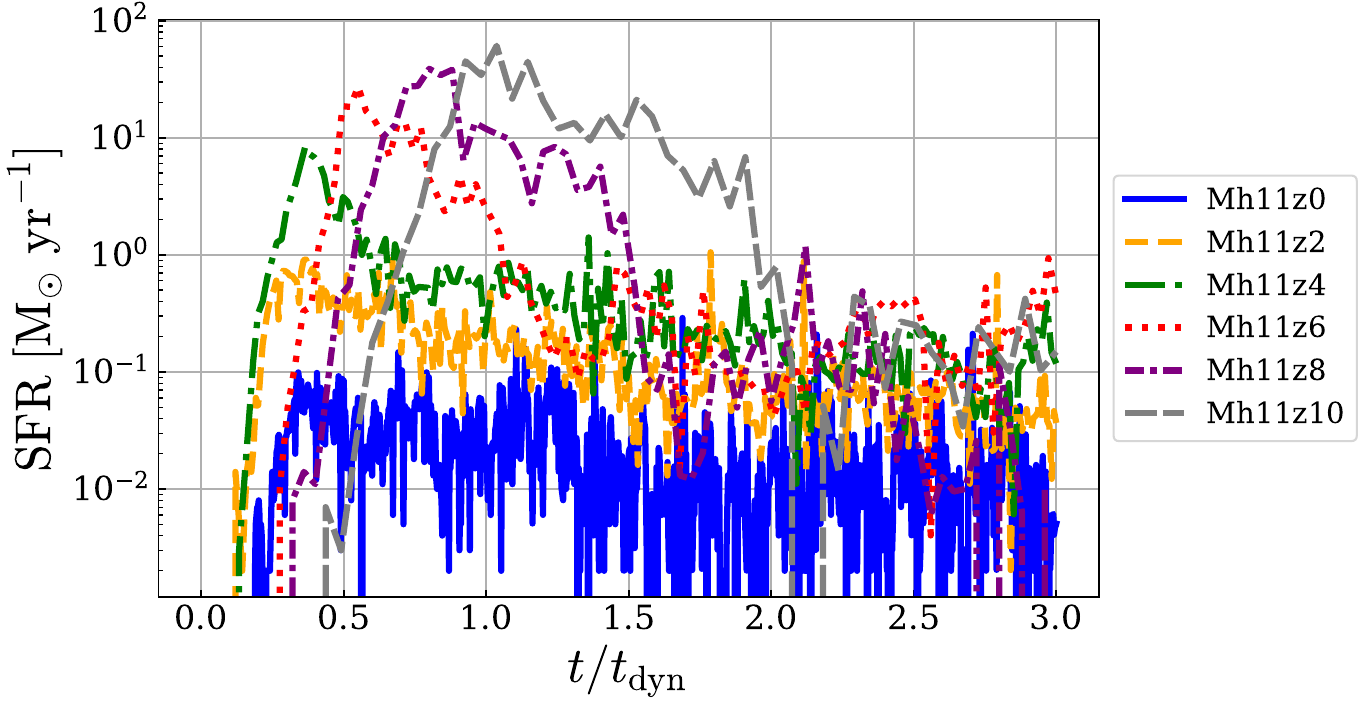}\\[2ex]
    \includegraphics[width=\columnwidth]{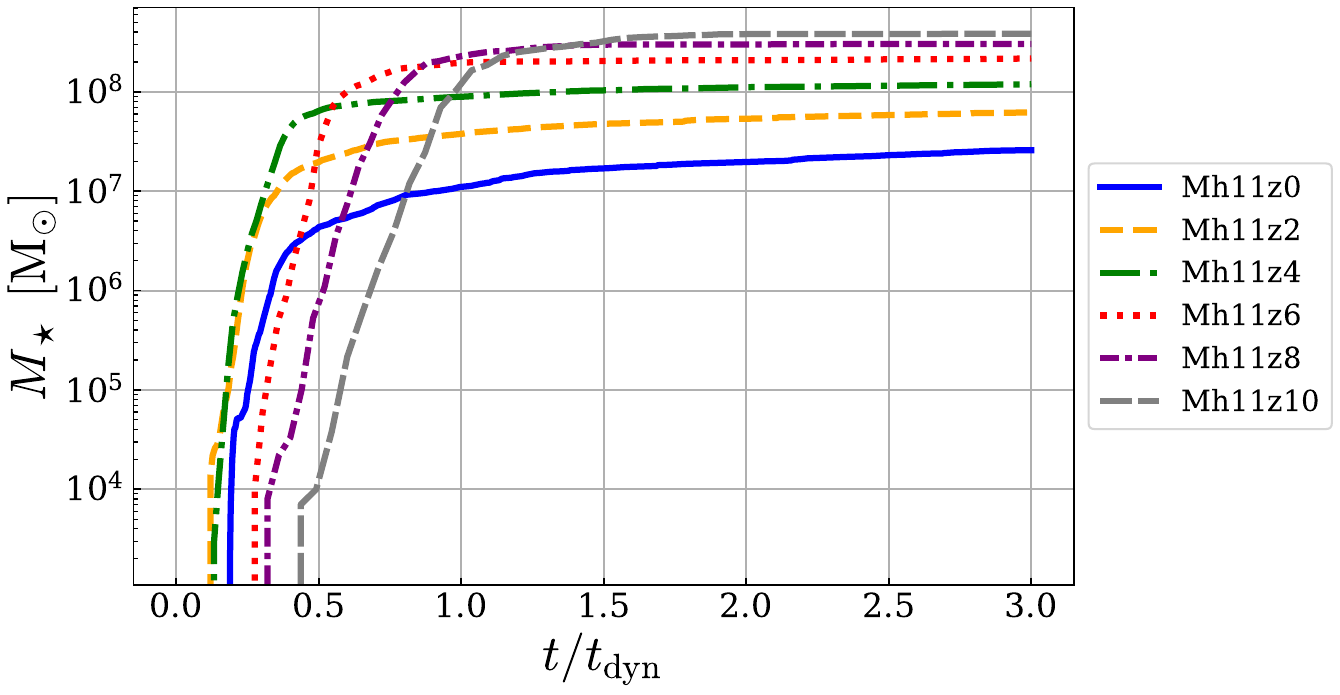}
    \caption{Same as Fig.~\ref{fig:sfh_Mh10_CSF}, but for the $10^{11}\,\Msun$ halo simulations.}
    \label{fig:sfh_Mh11_CSF}
\end{figure}

The star formation histories of the $10^{11}\,\Msun$ halo simulations are shown in the top panel of Fig.~\ref{fig:sfh_Mh11_CSF}.
All simulations exhibit continuous star formation activity, even in the $z=0$ case.
However, the $z=4$, $6$, $8$, and $10$ runs experience pronounced bursty phases, after which the SFRs decline by several orders of magnitude.
Up to $t\approx t_\mathrm{dyn}$, the high-$z$ galaxies show strong SFR peaks whose amplitudes increase with galactic compactness.
The timescale from the SFR peak to the subsequent rapid decline is typically $\sim0.5$--$1\,t_\mathrm{dyn}$, corresponding to a few tens of Myr.

The bottom panel of Fig.~\ref{fig:sfh_Mh11_CSF} shows the evolution of the total stellar mass formed over three rotational periods.
As in the $10^{10}\,\Msun$ cases, most stars are formed until $t\approx t_\mathrm{dyn}$.
The resulting galactic SFEs are $\approx2\times10^{-2}$, $7\times10^{-2}$, $1\times10^{-1}$, $2\times10^{-1}$, $3\times10^{-1}$, and $4\times10^{-1}$ for $z=0$, $2$, $4$, $6$, $8$, and $10$, respectively.
These results suggest that galaxies embedded in $10^{11}\,\Msun$ haloes can form stars several times more efficiently than those in $10^{10}\,\Msun$ haloes.

\subsubsection{Bound star clusters}
\label{results:halo_mass:clusters}

\begin{figure}
    \centering
    \includegraphics[width=\columnwidth]{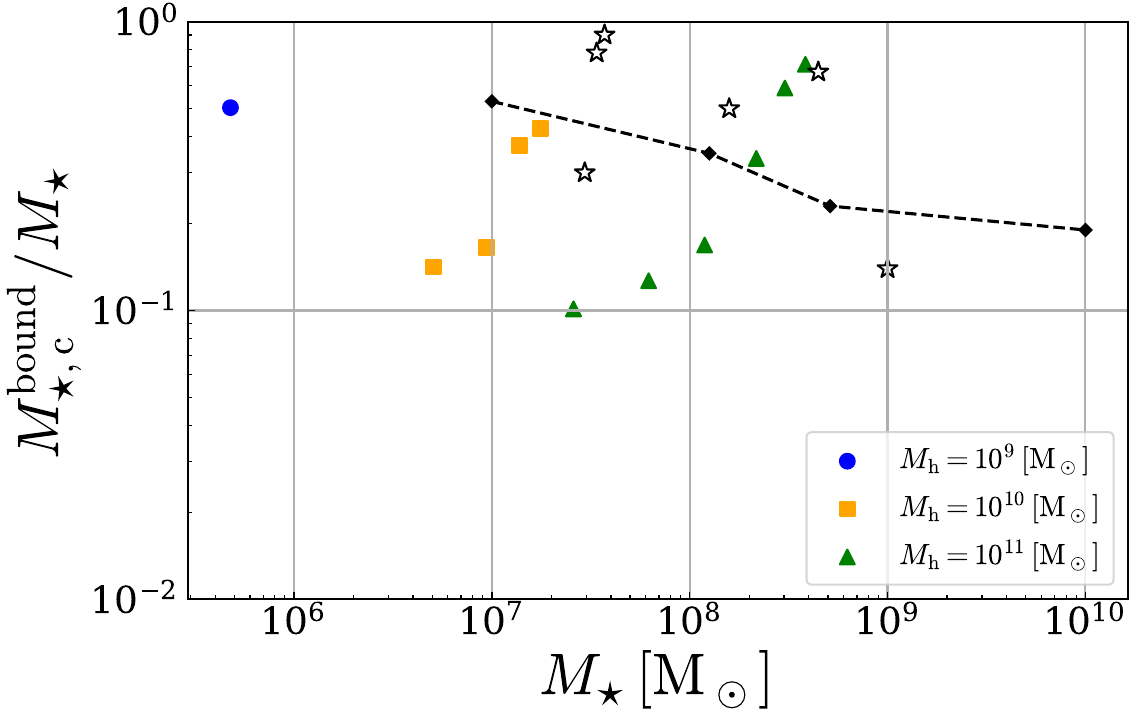}
    \caption{The fraction of bound star cluster mass relative to the total stellar mass in our simulations and observational results. Blue circles, orange squares, and green triangles represent our simulation results for $M_\mathrm{h}=10^9$, $10^{10}$, and $10^{11}\,\Msun$ haloes at $t=3t_\mathrm{dyn}$, respectively. White star markers show observational estimates for high-$z$ clumpy galaxies, including BulletArc-z11 \citep[$z=11.10$,][]{bradac_2025}, Cosmic Gems arc \citep[$z=9.63$,][]{adamo_2024,bradley_2025,messa_2026,vanzella_2026}, Cosmic Grapes \citep[$z=6.07$,][]{fujimoto_2025}, the Sunburst galaxy \citep[$z=2.37$,][]{vanzella_2022}, and Misty Moons \citep[$z\sim11$--$12$,][]{nakane_2025}. The median values of observed clumpy galaxies in the redshift range $1<z<5.5$ are shown as black diamonds connected by a dashed line \citep[][]{claeyssens_2025}. We note that the observational data represent the fraction of cluster mass, which may not necessarily correspond to gravitationally bound mass.}
    \label{fig:bound_cluster_cmp}
\end{figure}

We next compare the properties of bound star clusters among simulations with different halo masses.
Figure~\ref{fig:bound_cluster_cmp} shows the bound cluster mass fraction as a function of total stellar mass at $t=3t_\mathrm{dyn}$ for the $M_\mathrm{h}=10^{9}, 10^{10}$ and $10^{11}\,\Msun$ simulations.
In the $M_\mathrm{h}=10^{9}\,\Msun$ simulations, only the $z=10$ case forms self-gravitationally bound star clusters, although the other cases produce stars as shown in Fig.~\ref{fig:sfh_Mh9_CSF}.
The bound fraction reaches $\sim50\,\%$ at $t=3t_\mathrm{dyn}$, comparable to the value obtained in the $z=10$ case of the $M_\mathrm{h}=10^{10}\,\Msun$ simulation.

While self-gravitationally bound star clusters form only in limited environments in the $10^{9}\,\Msun$ halo simulations, they are produced in all of the $10^{11}\,\Msun$ halo runs.
The dependence of the fraction on stellar mass is more clearly seen in the $M_\mathrm{h}=10^{11}\,\Msun$ simulations than in the $M_\mathrm{h}=10^{10}\,\Msun$ cases.
The stellar masses span a range of $M_\star\approx 3\times10^7$--$4\times10^8\,\Msun$, and the bound cluster fractions systematically increase with increasing stellar mass.
This trend is broadly consistent with the results obtained for the other halo mass cases.
However, galactic compactness also plays an important role in determining the bound cluster fraction.
For example, in the $10^{9}\,\Msun$ halo simulation at $z=10$, the total stellar mass is only $M_\star\approx5\times10^{5}\,\Msun$, while the bound cluster fraction reaches $\approx50\,\%$.
This value is significantly higher than expected from the relation found in the $10^{10}\,\Msun$ and $10^{11}\,\Msun$ halo simulations.
Therefore, the correlation between stellar mass and bound cluster fraction appears to hold primarily when the halo mass is fixed, and only the galactic compactness varies.

We here compare our simulation results with observational data. 
Figure~\ref{fig:bound_cluster_cmp} also shows the fraction of bound cluster mass relative to the total stellar mass observed in high-$z$ clumpy galaxies \citep[][]{bradac_2025,adamo_2024,bradley_2025,messa_2026,vanzella_2026,fujimoto_2025,vanzella_2022,nakane_2025,claeyssens_2025}. 
We note that, while we use bound clusters in our simulations, the observational estimates are based on cluster masses, and therefore the two quantities may not be directly comparable.
This may introduce systematic differences in the inferred fractions between simulations and observations.
We find that the stellar mass range in our $10^{10}$ and $10^{11}\,\Msun$ halo simulations is broadly consistent with the observed sample, whereas the $10^{9}\,\Msun$ halo case is lower by a factor of $\sim 10$ compared to observations.
On the other hand, the runs that form bound clusters yield fractions of
$\approx 10$--$70$ per cent, which is consistent with the observed range.
Although our simulations do not reproduce the most extreme systems, such as the Misty Moons \citep[][corresponding to $M_\star \approx 3\times10^{7}\,\Msun$ and cluster mass fractions of $\gtrsim 80\,\%$]{nakane_2025}, several of our results are consistent with the median trend of observed clumpy galaxies (shown as black diamonds connected by a dashed line). 
These results suggest that our star formation model may provide a viable explanation for the formation of observed clumpy galaxies.
However, our current galaxy models are simplified and non-cosmological.
To fully investigate the formation pathways of high-$z$ clumpy galaxies, cosmological simulations incorporating our star formation model will be required.

\section{Discussion}
\label{discussion}

\subsection{Dependence on star formation models}
\label{discussion:sf_models}

\begin{figure}
    \includegraphics[width=\columnwidth]{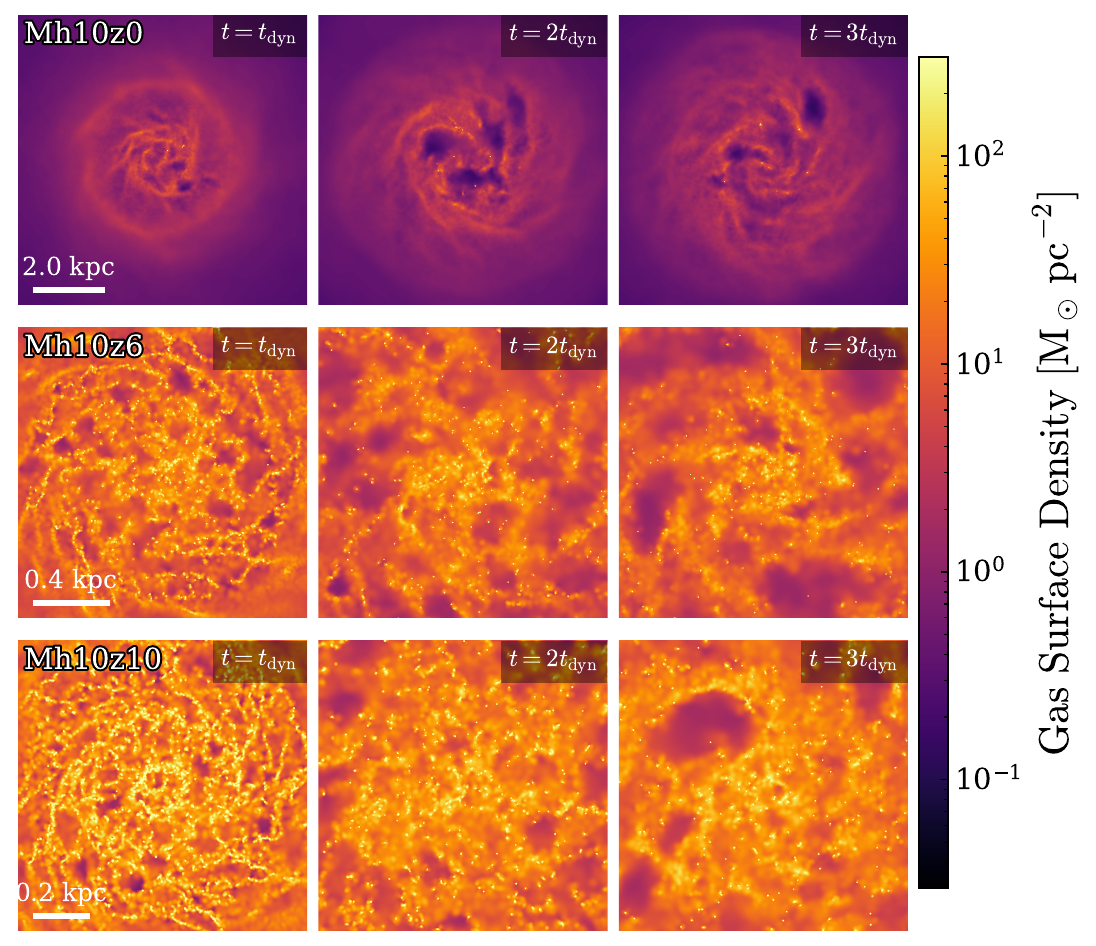}\\[2ex]
    \includegraphics[width=\columnwidth]{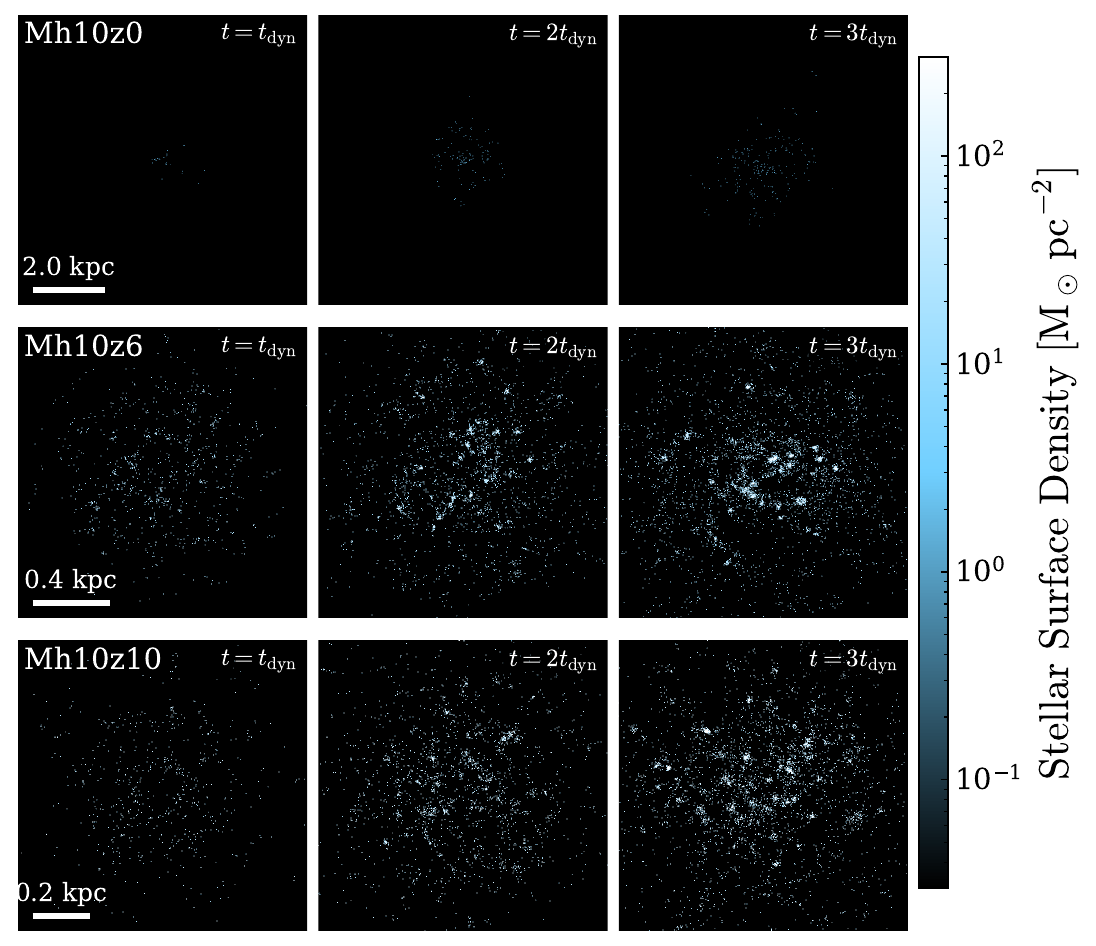}
    \caption{Same as Fig.~\ref{fig:surfden_Mh9_CSF}, but for the $10^{10}\,\Msun$ halo simulations using the KS model.}
    \label{fig:surfden_Mh10_KS}
\end{figure}

For comparison with the simulations using the CSF model, we present the results of the simulations employing the KS model.
The galactic morphologies and star formation activity in the KS simulations differ significantly from those in the CSF runs.
As shown in the top panel of Fig.~\ref{fig:surfden_Mh10_KS}, gaseous disc structures are maintained over three rotational periods in all simulations.
Unlike the CSF cases, even the $z=10$ simulation does not completely expel its gas, although hot bubbles with sizes of several hundred $\pc$ are generated.
In addition, the stellar distributions shown in the bottom panel of Fig.~\ref{fig:surfden_Mh10_KS} do not exhibit spiral structures as clearly as those seen in the CSF simulations.
These results suggest that the choice of star formation prescription is a key factor governing the morphological evolution of galaxies.

\begin{figure}
    \centering
    \includegraphics[width=\columnwidth]{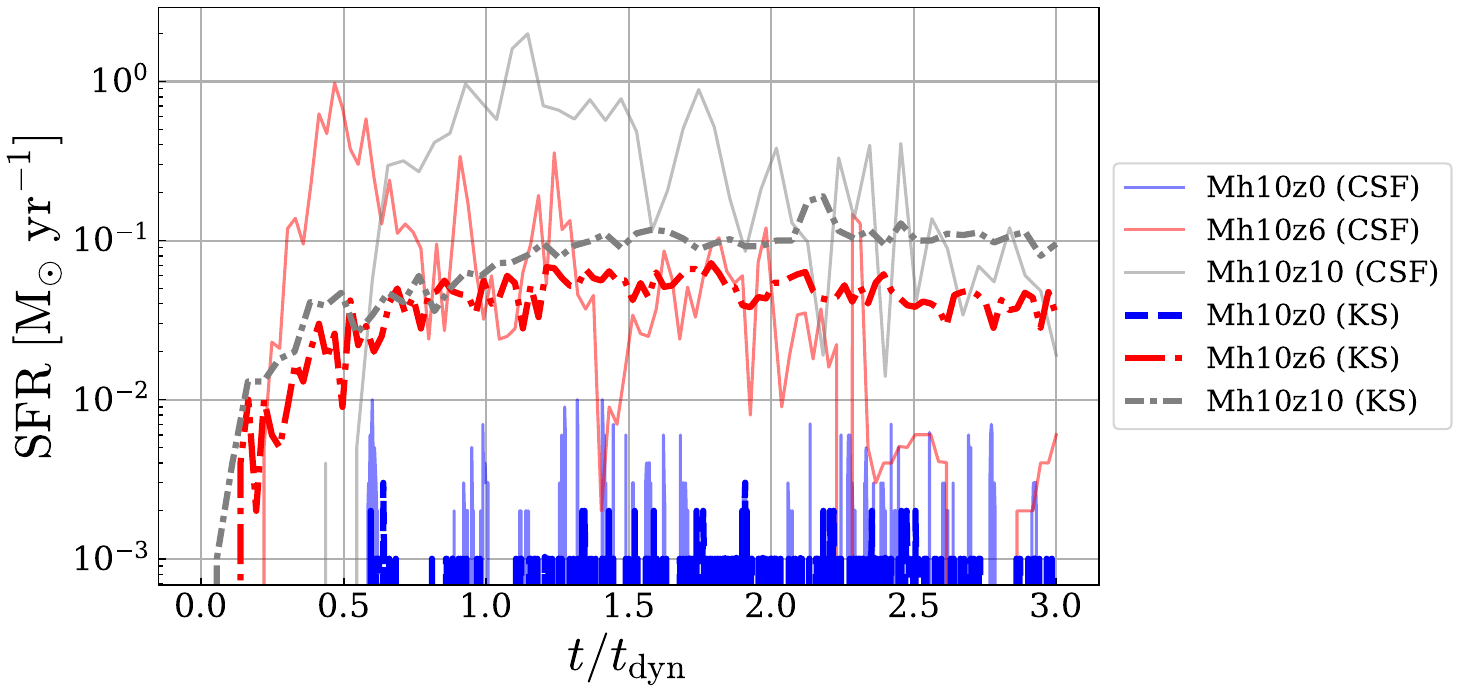}\\[2ex]
    \includegraphics[width=\columnwidth]{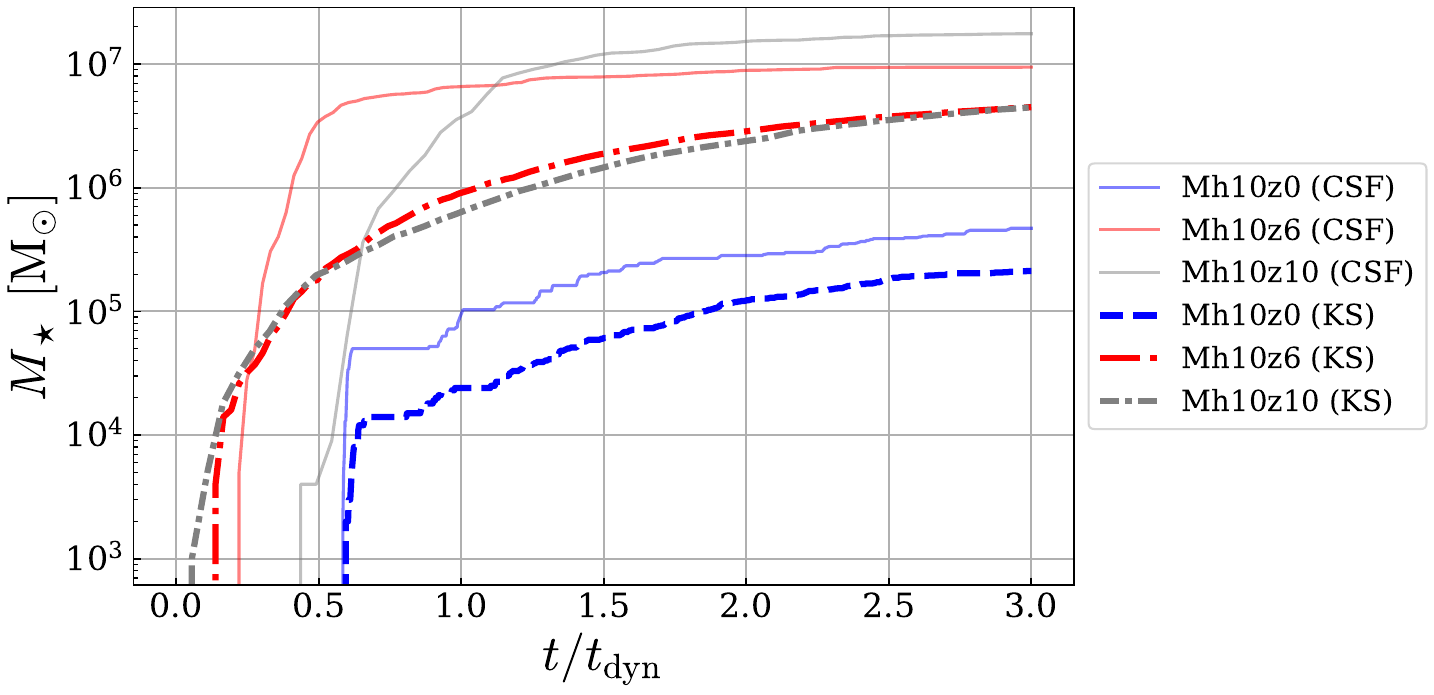}
    \caption{Comparison of SFRs (top) and total stellar masses (bottom) between the CSF and KS simulations for haloes with $M_\mathrm{h}=10^{10}\,\Msun$ at $z=0$, 6, and 10. The thin, translucent lines show the results of the CSF simulations, which are the same as those shown in Fig.~\ref{fig:sfh_Mh10_CSF}, while the thick lines represent the results of the KS simulations.}
    \label{fig:sfh_Mh10_KS}
\end{figure}

Star formation activity in the KS simulations is generally weaker than in the CSF simulations, as shown in the top panel of Fig.~\ref{fig:sfh_Mh10_KS}.
In the $z=0$ simulation with the KS model, star formation is only marginally resolved because the SFR remains close to the minimum resolvable value in our simulations, $\sim10^{-3}\,\sfrunit$.
The other KS simulations exhibit continuous star formation over three rotational periods, but unlike the CSF runs, they do not show strongly bursty star formation histories.
Overall, the SFRs tend to increase with galactic compactness, namely towards higher $z$.
However, even in the $z=10$ simulation with the KS model, the SFR reaches at most $\sim10^{-1}\,\sfrunit$, which is approximately one order of magnitude lower than in the corresponding CSF simulation.
Because star formation proceeds continuously in the KS simulations, the total stellar mass gradually increases even after $t=t_\mathrm{dyn}$.
In the KS simulations using $M_\mathrm{h}=10^{10}\,\Msun$ haloes, the galactic SFEs at $t=3t_\mathrm{dyn}$ are $\mathrm{SFE}\approx2\times10^{-3}$, $1\times10^{-2}$, $3\times10^{-2}$, $5\times10^{-2}$, $5\times10^{-2}$, and $5\times10^{-2}$ for $z=0$, $2$, $4$, $6$, $8$, and $10$, respectively.
Therefore, the KS model results in systematically less efficient star formation than the CSF model by factors of a few.

\subsection{Implications for cosmological simulations}
\label{discussion:appl}

The KS prescription ties the star formation rate of a gas particle to the empirical relation between SFR and gas surface densities \citep{kennicutt_1998}, and this model is insensitive to numerical resolution.
Cosmological simulations adopting it reproduce several statistical properties of galaxies, including the cosmic star formation history and the star-forming main sequence \citep[e.g.][]{ schaye_2015_eagle,yajima_2022}, but tend to under-produce the massive quenched galaxies already observed at $z \gtrsim 2$ \citep[e.g.][]{daddi_2005,tacchella_2015,tanaka_2019,esdaile_2021,tanaka_2024}.
Our results are consistent with this tendency.
The KS runs form stars continuously even in the most compact discs, whereas the CSF runs exhibit intermittent, bursty, and subsequently quenched star formation histories.

This contrast leads to a testable prediction.
In a cosmological setting, the CSF model should produce a larger scatter in the star-forming main sequence, and a larger population of temporarily quenched systems, than KS-based models such as FOREVER22.
We caution, however, that the CSF model relies on the on-the-fly identification of gas clumps and is therefore sensitive to the gas mass resolution, so that the amplitude of the predicted burstiness should be verified against resolution before it is compared with observations.
We will address these points, and the impact of the CSF model on galactic properties more generally, in forthcoming cosmological simulations.

\section{Conclusion}
\label{conclusion}

We have investigated how the compactness of galactic discs regulates clumpy star formation and the formation of bound star clusters.
To this end, we have developed a new star formation model (CSF) in which star-forming gas clumps are identified on-the-fly and converted into stars with an efficiency $\epsilon_\mathrm{SF}$ set by the clump surface density $\Sigma_\mathrm{c}$, calibrated against the radiation hydrodynamics simulations of \citet{fukushima_2021}.
We apply the model to isolated disc galaxies embedded in haloes of $M_\mathrm{h} = 10^{9}$, $10^{10}$ and $10^{11}\,\Msun$ with a fixed gas-to-halo mass ratio of $0.01$, adopting disc scale radii appropriate for $z = 0$--$10$ as a proxy for compactness.
Our main results are summarised as follows.

\begin{enumerate}
    \item Compact discs form more massive and denser gas clumps. In the $10^{10}\,\Msun$ haloes, the maximum clump mass increases from $\sim 10^{5.1}$ to $10^{7.4}\,\Msun$ and the maximum clump surface density from $\sim 10^{2.9}$ to $> 10^{4.0}\,\Msun\,\mathrm{pc}^{-2}$ between $z = 0$ and $10$. The fraction of clumps exceeding $\Sigma_\mathrm{c,th} = 300\,\Msun\,\mathrm{pc}^{-2}$ rises from $0.25$ at $z = 0$ to $0.34$ at $z = 6$ and $0.38$ at $z = 10$, and the most compact discs host clumps with $\epsilon_\mathrm{SF} \sim 1$.

    \item This shift in clump properties translates directly into more efficient and burstier galactic star formation. In the $10^{10}\,\Msun$ haloes, the global SFE at $t = 3t_\mathrm{dyn}$ increases monotonically from $5 \times 10^{-3}$ at $z = 0$ to $2 \times 10^{-1}$ at $z = 10$, with peak SFRs of $\sim 1\,\Msun\,\mathrm{yr}^{-1}$ reached before $t \approx t_\mathrm{dyn}$ for $z \geq 6$. Less compact discs instead form stars intermittently at $10^{-2}$--$10^{-1}\,\Msun\,\mathrm{yr}^{-1}$.

    \item The accompanying feedback disrupts the gaseous discs of the most compact systems. The $z \geq 8$ galaxies retain little gas near the centre by $t = 3t_\mathrm{dyn}$, whereas the $z \leq 2$ discs survive three rotations. Stellar spiral arms show the opposite trend, becoming progressively more prominent towards higher $z$, indicating that compact discs assemble dynamically coherent stellar structures on timescales of $\sim 100$ Myr even while their gas is being expelled.

    \item Self-gravitationally bound star clusters form only in compact discs. No bound clusters appear at $z \leq 2$ in the $10^{10}\,\Msun$ haloes, whereas the bound cluster mass fraction reaches $\approx 40$ per cent at $z \geq 8$. The threshold compactness depends on halo mass. In the $10^{9}\,\Msun$ haloes only the $z = 10$ run forms bound clusters, while in the $10^{11}\,\Msun$ haloes every run does, with fractions rising from $\approx 10$ per cent at $z = 0$ to $\approx 70$ per cent at $z = 10$. Halo mass therefore acts alongside compactness in regulating bound cluster formation, and the global SFE spans $2 \times 10^{-3}$--$5 \times 10^{-2}$, $5 \times 10^{-3}$--$2 \times 10^{-1}$ and $2 \times 10^{-2}$--$4 \times 10^{-1}$ for $M_\mathrm{h} = 10^{9}$, $10^{10}$ and $10^{11}\,\Msun$, respectively.

    \item The choice of star formation prescription is itself a leading factor. In the $10^{10}\,\Msun$ haloes, the KS runs retain their gas discs at all redshifts, show no strongly bursty histories, reach peak SFRs an order of magnitude below the CSF runs at $z = 10$, and saturate at SFE $\approx 5 \times 10^{-2}$ for $z \geq 6$ instead of continuing to rise with compactness. The resulting global SFEs are lower than in the CSF runs by factors of a few.

    \item Among the runs that form bound clusters, the bound cluster mass fractions at $t = 3t_\mathrm{dyn}$ range from $\approx 10$ to $\approx 70$ per cent, consistent with the cluster mass fractions inferred for observed high-$z$ clumpy galaxies, and the stellar masses of our $10^{10}$ and $10^{11}\,\Msun$ haloes overlap the observed sample. Our model thus offers a viable route to forming such systems, although it does not reach the most extreme fractions of $\gtrsim 80$ per cent reported for the Misty Moons \citep{nakane_2025}.
\end{enumerate}

We emphasise that these simulations are idealised since the discs are isolated and non-cosmological, the halo-to-disc mass ratio is fixed, and no stellar disc is included in the initial conditions.
Thus, environmental effects, gas accretion, and hierarchical growth are absent by construction.
The controlled setup adopted here allows the effects of compactness to be isolated, and the next step is to test the same physics in a cosmological context, where disc sizes, gas accretion and merger histories are set self-consistently rather than as free parameters.
We will carry out such simulations with the CSF model in the near future.

\section*{Acknowledgements}

HY acknowledges the funding support for this work from MEXT/JSPS KAKENHI 21H04489, 26H02061, JST FOREST Program, Grant Number JP-MJFR202Z. 
Numerical computations were carried out on Cray XD2000 at the Center for Computational Astrophysics, National Astronomical Observatory of Japan.
We thank the developers of the following Python analysis tools:
\textsc{matplotlib} \citep[][]{hunter_2007_matplotlib}, \textsc{numpy} \citep[][]{vanderwalt_2011_numpy}, and \textsc{yt} \citep[][]{turk_2011_yt}.

\section*{Data Availability}

The data underlying this article can be shared on a reasonable request to the corresponding author.



\bibliographystyle{mnras}
\bibliography{references}




\appendix




\bsp	
\label{lastpage}
\end{document}